\documentclass{sig-alternate}
\usepackage{graphicx}
\usepackage{caption}
\usepackage{subcaption}

\makeatletter
 \let\@copyrightspace\relax
\makeatother
\usepackage{balance}  
\usepackage[linesnumbered, ruled]{algorithm2e}
\usepackage{color}

\begin{document}
\newdef{lemma}{Lemma}
\newdef{proposition}{Proposition}
\newdef{theorem}{Theorem}
\newdef{conclusion}{Conclusion}
\renewcommand{\algorithmcfname}{Algorithm}

\title{On Order-independent Semantics of the Similarity Group-By Relational Database Operator}

\newcommand\Mark[1]{\textsuperscript#1}
\numberofauthors{1}

\author{
Mingjie Tang\Mark{1}$\;$ $\;$  Ruby Y. Tahboub\Mark{1}$\;$ $\;$ Walid G. Aref\Mark{1}$\;$ $\;$ Qutaibah M. Malluhi\Mark{2}$\;$ $\;$ Mourad Ouzzani\Mark{3}
\\
\affaddr{\Mark{1}Purdue University $\;$ \Mark{2}Qatar University $\;$ \Mark{3}Qatar Computing Research Institute}
\\
\email{\{tang49,rtahboub,aref\}@cs.purdue.edu, qmalluhi@qu.edu.qa, mouzzani@qf.org.qa}
}

\maketitle

\begin{abstract}
Similarity group-by (SGB, for short) has been proposed as a relational database operator
to match the needs of emerging database applications. Many SGB operators that extend SQL
have been proposed in the literature, e.g., similarity operators in the one-dimensional
space. These operators have various semantics. Depending on how these operators are
implemented, some of the implementations may lead to different groupings of the data.
Hence, if SQL code is ported from one database system to another, it is not guaranteed
that the code will produce the same results.
In this paper, we investigate the various
semantics for the relational similarity group-by operators in the multi-dimensional space.
We define the class of order-independent SGB operators that produce the same results
regardless of the order in which the input data is presented to them.
Using the notion of interval graphs borrowed from graph theory, we prove that, for
certain SGB operators, there exist order-independent implementations.
For each of these operators, we provide a sample algorithm that is order-independent.
Also, we prove
that for other SGB operators, there does not exist an
order-independent implementation for them, and hence these SGB operators are ill-defined
and should not be adopted in extensions to SQL to realize similarity group-by. In this
paper, we introduce an SGB operator, namely SGB-All, for grouping multi-dimensional
data using similarity.
SGB-All forms groups such that a data item, say O, belongs to a group, say G, if and only
if O is within a user-defined threshold from all other data items in G.
In other words, each group in SGB-All forms a clique of nearby data items in the multi-dimensional
space.
One case
that arises in both SGB operators is when a data item qualifies the membership condition of multiple
groups. For example, in the case of SGB-All, a data item, say O, can form a clique with two groups,
say $G_1$ and $G_2$. We propose three semantics for handling data items that overlap multiple groups,
namely, Eliminate, Duplicate, and New-Group. We prove that all three options are order-independent,
i.e., there is at least one algorithm for each option that is independent of the presentation order
of the input data.
\end{abstract}


\section{Introduction}
\begin{sloppypar}
Forming groups of data items to support decision making is a fundamental function of a
database management system.
Traditionally, grouping is performed by aggregating the tuples with equal values on a
certain subset of attributes into the same groups. However, in some application domains,
e.g., business intelligence, sensor networks, and location-based queries,
users are often interested in grouping based on similar rather than strictly equal values.
\end{sloppypar}

\begin{sloppypar}
Clustering is one way of grouping similar and closeby data items together.
Clustering is a well-known operation in data mining and machine learning with well-established tools, e.g., Weka~\cite{EEEexample:hall09}.
In most cases, clustering is performed outside the database system, which leads to several issues:
First, it creates a costly impedance mismatch
that results from having to extract the data outside of the database to perform the clustering.
Moreover, based on the needs of the underlying applications, the output clusters may need to be further processed by SQL to
filter out some of the
clusters and perform further SQL operations on the remaining ones.
Hence, it is of great benefit to develop  a practical and fast
similarity group-by operator that can be embedded within a SQL query
and that is compatible with other SQL operators.
This would allow answering complex similarity-based queries efficiently.
\end{sloppypar}

\begin{sloppypar}
Silva et al.~\cite{IEEEexample:silva2009similarity,VLDB:SilvaALPA13} introduce the
similarity group-by operator (SGB) in a variety of flavors that can be used in an SQL query in
combination with other relational operators and that can take advantage of database query
optimization techniques, e.g., pushing joins under similarity-group-by.
However, the problem with the SGB operators in~\cite{IEEEexample:silva2009similarity,VLDB:SilvaALPA13} is two-fold.
First, these SGB operators are one-dimensional, i.e., they group
the data of each attribute independently from the other attributes.
Hence, they cannot detect correlations among the various attributes.
Secondly, the semantics of these SGB operators such that these operators are order-dependent, i.e.,  the outcome of these SGB operators may differ depending on the order in which the input tuples are processed.
Therefore, if SQL code is ported from one database system to another, there is no guarantee that it will produce the same results.
\end{sloppypar}

\begin{sloppypar}
In this paper, we investigate the semantics of various multi-attribute
similarity group-by operators. We define the class of
multi-dimensional order-independent SGB operators that will produce the same results regardless of the
order in which the input tuples are being processed.
We demonstrate that, for several of these SGB operators, there exist order-independent implementations for them.
\end{sloppypar}

\begin{sloppypar}
In contrast to
relational group-by where a tuple may belong to one and only one group, in similarity
group-by, a tuple may belong to multiple groups if that tuple is in the proximity of these
groups.
We provide several
similarity group-by semantics to handle these overlapping tuples
and prove that the algorithms realizing these semantics are order-independent.
\end{sloppypar}

\begin{sloppypar}
The contributions of this paper can be summarized as follows.
\begin{enumerate}

\item We introduce the notion of order-independent similarity group-by operators (SGB)
and provide several definitions of SGB operators that are order-independent.
We give proofs
that several of these  SGB operators are actually order-independent.
In contrast, for other
SGB operators that are not order-in\-depend\-ent, we give  counter-examples that illustrate
their order-dependence.

\item We study the cases when a tuple overlaps more than one group.
We provide several semantics to handle overlapping tuples
and prove that these semantics are order-independent.

\item
We provide sample order-independent algorithms for the introduced SGB operators to minimize the running time complexity via group bound, then
analyze the proposed algorithm's complexity and order independent properties.
\end{enumerate}
\end{sloppypar}

\begin{sloppypar}
The rest of this paper proceeds as follows.
Section~\ref{sec-motivation} motivates the need for order-independent SGB operators.
Section~\ref{sec-related} presents and discusses related work.
Section~\ref{sec-defs} provides relevant background material.
Section~\ref{sec-SGB-ops} presents existing and new SGB operators and investigates their order-independence.
Section~\ref{sgb-alg} gives efficient algorithms to implement the SGB operators along with analyses of their complexities.
Section~\ref{sec-proofs} presents the proofs of correctness for order-independence for the new SGB operators based on the proposed group bound approach using the notion of interval graphs.
Finally, Section~\ref{sec-con} contains concluding remarks.
\end{sloppypar}

\section{Motivation}
\label{sec-motivation}

\begin{figure}
\centering
\includegraphics[width=2.9in,height=2.1in]{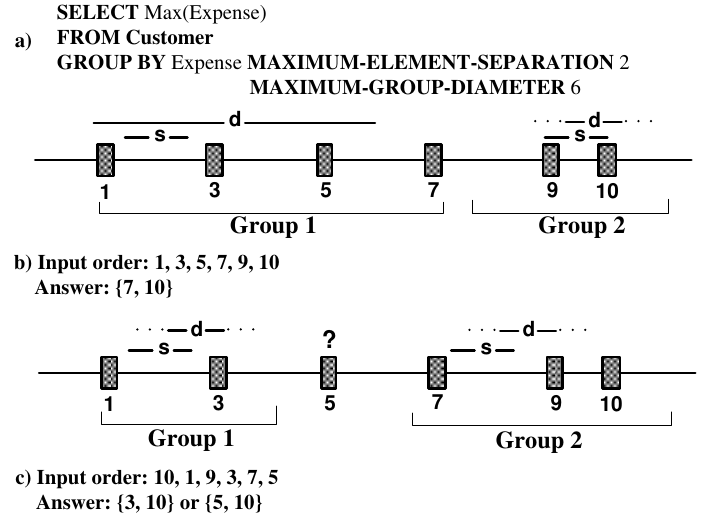}
\caption{Unsupervised Similarity Group-by SGB-U}
\label{figure:SGB-U}
\end{figure}

\begin{sloppypar}
To illustrate the need for order-independent similarity group-by operators, consider the
unsupervised similarity group-by operator (SGB-U), introduced in~\cite{IEEEexample:silva2009similarity}.
SGB-U is a one-dimensional SGB operator that takes as input one attribute/column of the input table and extends SQL's standard
Group-by with two similarity predicates:
(i)~The \textit{maximum$\_$element$\_$separation} (or {\em separation}) defines the maximum distance between two adjacent elements in a
group.
It controls the compactness or closeness among elements within the same group.
(ii)~The \textit{maximum$\_$group$\_$diameter} (or {\em diameter}) is the distance between the two farthest elements within a group.
For example, in the linear space, a group with elements $\left\lbrace1,5,6\right\rbrace$ has
a {\em separation} of 4 and a {\em diameter} of 5.
\end{sloppypar}

\begin{sloppypar}
Consider the query example illustrated in Figure~\ref{figure:SGB-U}a that uses SGB-U.
Given the relation $customer$, the query finds the maximum expense of groups of customers with similar expense values.
Similarity in this query is defined by a separation value of 2 and a diameter value of 6.
Suppose the values of the \textit{expense} attribute in Table~\textit{Customer} are processed in the following order:
$\left\lbrace  1,3,5,7,9,10 \right\rbrace$.
In this case, elements  $\left\lbrace  1,3,5,7\right\rbrace$ and elements $\left\lbrace 9,10 \right\rbrace$
form two distinct groups, Group~1 and Group~2, respectively.
For Group~1, the separation between adjacent elements
is within threshold 2 and the diameter of the group is 6.
Similarly, for Group~2, both the separation and diameter are equal to 1, as illustrated in Figure~\ref{figure:SGB-U}b.
Now, assume that SGB-U is presented with the input tuples in a different order, e.g., in the order
$\left\lbrace  10,1,9,3,7,5 \right\rbrace$ (refer to Figure\ref{figure:SGB-U}c).
In this case, two totally different groups are formed, namely
Group~1' with elements $\left\lbrace  1,3 \right\rbrace$ and Group~2' with elements $\left\lbrace  7,9,10 \right\rbrace$.
In addition, when processing the tuple with value~5,
it can be
inserted into both groups since~5 is within 2~units from the tuple with value~3 in Group~1 and the tuple with value~7 in Group~2.
Therefore, the outcome of SGB-U is order-dependent and overlaping data elements do not have a clear semantics.
As a result, given two implementations of SGB-U in two database systems, it is possible that the same SQL code when operating on
the same input database will produce completely different results.
The target of this paper is to identify multi-dimensinoal order-independent SGB operators that would produce the same result
regardless of the
presentation order of the input tables.
\end{sloppypar}

\section{Related Work}
\label{sec-related}
\begin{sloppypar}
In this section, we review the related work in three main areas:
clustering, existing similarity-based grouping operators in relational database systems,
and graph theory.

Clustering (also referred to as unsupervised learning) is a well-studied problem
in data mining. Clustering forms
groups of similar data items for the purpose of learning hidden knowledge about the data.
Clustering algorithms have been studied extensively,
e.g., see~\cite{IEEEexample:berkhin2006survey,IEEEexample:han2006data}.
Among the widely used clustering algorithms are the
K-means~\cite{IEEEexample:kanungo2002efficient}, disk-optimized algorithms, e.g.,
BIRCH~\cite{EEEexample:zhang1996birch},
and DBSCAN~\cite{IEEEexample:ester1996density}.
The key difference between SGB operators and clustering are:
(1)~SGB is a relational database operator that is integrated within relational query
evaluation pipelines with various grouping semantics. Hence, it avoids the impedance mismatch
experienced by standalong clustering packages that mandate extracting the data to be clustered out of the dbms.
(2)~In contrast to standalone clustering algorithms, SGB can be interleaved with other
relational operators, e.g., joins and selects,
(3)~standard relational query optimizations that apply to group-by are also applicable
to SGB, as illustrated in~\cite{IEEEexample:silva2009similarity,VLDB:SilvaALPA13}.
This is not feasible with
standalone clustering algorithms. Also, improved performance
can be gained by using database access methods that process multi-dimensional data.

The work in~\cite{IEEEexample:anderberg1973cluster} introduces proximity
semantics to overcome the limitation of distinct-value group-by operators.
Their proposed SQL construct, termed "ClusterBy", uses conventional clustering algorithms,
e.g., DBSCAN to realize similarity grouping. ClusterBy addresses the impedance mismatch issue
but does not address the order-independence issue that we address in this paper.

Previous work on similarity-based
grouping~\cite{IEEEexample:silva2009similarity,VLDB:SilvaALPA13}
introduce generic semantics for three core database SGB operators.
The unsupervised similarity group-by operatoer, SGB-U, realizes groups by applying two similarity
predicates that evaluate the compactness and diameter of a group.
The SGB-Around operator is a classifier operator where it uses predefined guiding points
(referred to as center points) to form groups around these centers.
The SGB-Delimited operator uses a set of delimiting points to segment the input tuples
into groups.
In addition to being one-dimensional and not handling correlated multi-attributes,
these three operators overlook the order in which the input tuples are processed and its
effect on the operator's outcome.
This paper builds on~\cite{IEEEexample:silva2009similarity,VLDB:SilvaALPA13} with a focus
on order-independent SGB semantics for multi-dimensional data.

The SGB order-independence proofs given in this paper build on previous work on graph
theory including interval graphs~\cite{IEEEexample:mckee1999topics} and maximal
cliques~\cite{EEEexample:Karp72}
A clique is a connected subgraph, where each vertex has an edge to every other vertex in
the clique subgraph.
The maximal clique problem finds the maximum size clique within a graph and is an
extensively studied problem~\cite{EEEexample:Karp72}.
For multi-dimensional data, when allowing data tuples to belong to multiple groups, e.g.,
as in the case of tuples that overlap multiple groups, listing maximal cliques sets becomes exponential.
A large body of work addresses approximate solutions for the maximal clique
problem (e.g., see~\cite{Clique:Bomze99themaximum}).
\cite{SIGMOD13:triangleProblem}~approximates the clique problem for disk-based data sets
by finding all three-vertex cliques (i.e., triangles) in the graph.
A fundamental difference between graph search problems and the proposed SGB operators in this paper is
that we do not build any graph beforehand. Instead, the edges between the tuples are implicit and are formed dynamically
based on the similarity between the tuples (i.e., the vertexes), and the SGB operators incrementally process the input
and form the output groups.
\end{sloppypar}

\section{Background Material}
\label{sec-defs}

In this section, we provide background material and formally introduce the class of order-independent similarity group-by operators.


\newdef{definition}{Definition}

\begin{definition}
A \textbf{similarity measure} is defined by a distance function $\delta$ that uses the Minkowski distance
$L_p$~\cite{Mdistance:Kruskal1964}.

In this paper, we consider the following two Minkowski distances, where $t_x$ is a tuple and $t_{xy}$ is its $y^{th}$  attribute:
\begin{itemize}
	\item The Euclidean Distance:\\
    $L_2 : \delta_2(t_i, t_j) = \sqrt{\displaystyle\sum_{i'}\left(t_{ii'}-t_{ji'}\right)^2}$
		
	\item The Maximum infinity distance: \\
    $L_\infty : \delta_\infty(t_i, t_j) =
		\displaystyle \max_{i'}\left|t_{ii'}-t_{ji'}\right|$.
\end{itemize}	
\end{definition}

\begin{sloppypar}
\begin{definition}
A \textbf{similarity predicate} $\xi(\delta, \epsilon)$ is a Boolean expression that uses the distance
function $\delta$ and the threshold $\epsilon$, and returns TRUE
iff the distance between its two input tuples $t_i$ and $t_j$ is less than $\epsilon$, i.e.,
\[
	\xi_{\delta,\epsilon}(t_i, t_j) : \delta(t_i, t_j) \le \epsilon.
\]
In this case, the two tuples $t_i$ and $t_j$ are said to be similar.
\end{definition}
\end{sloppypar}


\begin{sloppypar}
\begin{definition}
Let $T$ be a relation that consists of a tuple set $\left\lbrace t_{1}, ..., t_{m} \right\rbrace $. Each tuple $t_i \in T$ is defined as
$
\left\lbrace {t_{i.}}_{GA_1}, ..., {t_{i.}}_{GA_d},  {t_{i.}}_{NGA_1}, ... {t_{i.}}_{NGA_k} \right\rbrace
$,
where the subset
$
GA_c = \left\lbrace {t_{i.}}_{GA_1} ... {t_{i.}}_{GA_d} \right\rbrace
$
represents a  multi-dimensional grouping attribute, and the subset
$
NGA = \left\lbrace {t_{i.}}_ { NGA_1} ... {t_{i.}}_{ NGA_k} \right\rbrace
$
represents the non-grouping attributes.
Let
$
{\xi}_{\delta,\epsilon}$ be similarity predicates. A \textbf{similarity Group-by operator}
$
{\mathcal{G}}_{\langle GA_c, ({\xi}_{\delta, \epsilon})\rangle} (T)
$
forms a set of answer groups $G_{s}$ by applying the similarity predicates ${\xi}_{\delta, \epsilon}$ on the elements of the multi-dimensional grouping attribute $GA_c$. That is, a pair of tuples, say $t_x$ and $t_y$, are in the same group iff:
$ \xi_{\delta, \epsilon}(t_x._{GAc}, t_y._{GAc}) = TRUE$.

\end{definition}
\end{sloppypar}




We study the effect of the order of presenting the input tuples of $T$ to the similarity group-by operator on
the outcome of the operator.
In particular, we define the necessary conditions
that guarantee that the operators' output groups $ G_s=\left\lbrace G_1, ..., G_m \right\rbrace $ are independent of any order of
the input tuples.
Based on these conditions, we then introduce and study new order-independent SGB operators.

\section{Similarity Group-By Operators}
\label{sec-SGB-ops}


In the section, we study the semantics of existing SGB operators as defined
in~\cite{IEEEexample:silva2009similarity} and formally prove whether or not they  belong to the class of order-independent SGB
operators. Then, we introduce several new SGB operators and prove their order-indendence and analyze their complexity.

\subsection{Unsupervised SGB (SGB-U)}
As illustrated by the example given in Section~\ref{sec-motivation}, SGB-U is order-dependent based on the counter-example given, and
hence the following lemma is true.

\begin{lemma}
The SGB-U operator is order-dependent.
\end{lemma}

Using the construction in Figure~\ref{figure:SGB-U}, permuting the order of input may produce different groupings.
The maximum diameter predicate determines the start- and end-points of a group. Hence, changing the order of processing
input may cause the start- and end-points of another group to shift resulting in a different grouping.

\begin{figure}
\centering
\includegraphics[width=2.9in,height=1.4in]{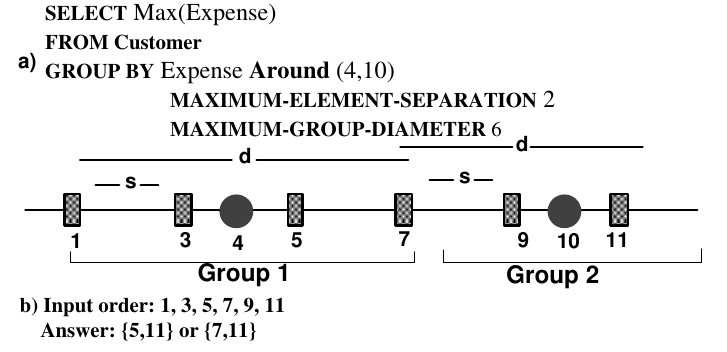}
\caption{Supervised Similarity Group-by Around SGB-A}
\label{figure:SGB-A}
\end{figure}

\subsection{SGB Around (SGB-A)}
\begin{sloppypar}
Another interesting SGB operator is the similarity group-by Around, or SGB-A for short~\cite{IEEEexample:silva2009similarity}.
SGB-A extends the SGB-U operator with a set of guiding points (termed central points),
where groups are composed around these points. Tuples of the input table are assigned to the closest center point.
The optional \textit{maximum$\_$element$\_$separation} predicate determines the maximum allowable distance between two adjacent
tuples in a group. The optional diameter predicate in SGB-A determines the group bounds, where the diameter is centered at the
guiding point (i.e., at position d/2).
\end{sloppypar}

Consider the query example illustrated in Figure~\ref{figure:SGB-A}a. Given the relation $customer$, the example query finds the
maximum expense of groups formed around center points~4 and~10. The seperation is set to the  value~2 and the maximum allowable
diameter is set to the value~6. Suppose that the input is processed in the following order
$\left\lbrace  1, 3, 5, 7, 9, 11 \right\rbrace$. In this case, Group~1 consists of $\left\lbrace  1, 3, 5 \right\rbrace$
while Group~2 consists of $\left\lbrace  9, 11 \right\rbrace$. The tuple with value~7 is within equal distance from both
center points. Hence, 7 can join either group. The semantics of SGB-A is ambiguous in the sense that it does not provide
arbitration semantics to resolve the case of overlapping tuples, i.e., ones that can belong to more than one group, and hence the
following lemma is true.

\begin{lemma}
The semantics of the SGB-A operator is ambiguous.
\end{lemma}

In general,
using a counter-example construction similar to the one in Figure~\ref{figure:SGB-A},
there may exists a tuple $e_i$ that is within equal distance from center points $c_1$ and $c_2$, and $c_1$  $\neq$ $c_2$.

\subsection{Order-independent SGB Operators}
\begin{sloppypar}
In this section, we introduce several order-independent SGB operators, namely the family of the SGB-All operators.
\end{sloppypar}

\begin{definition}
Given a table $T=\left\lbrace t_{1}, ..., t_{m}\right\rbrace$, where $t_{i}$ is a multi-dimensional tuple and $T'$ is a permutation of Table $T$. Let $GA$ be a multi-dimensional grouping attribute. The outcome of the \textbf{\textit{order-independent} Similarity Group-by} operator $\mathcal{\check{G}} $ is
independent of the order of the input tuples iff
\begin{equation*}
\mathcal{\check{G}}_{T_{GA}}(T) =  \mathcal{\check{G}}_{T'_{GA}}(T')
\end{equation*}
\end{definition}

\subsubsection{Group Definition}
\begin{figure}
        \centering
        \begin{subfigure}[b]{0.23\textwidth}
                \includegraphics[width=\columnwidth]{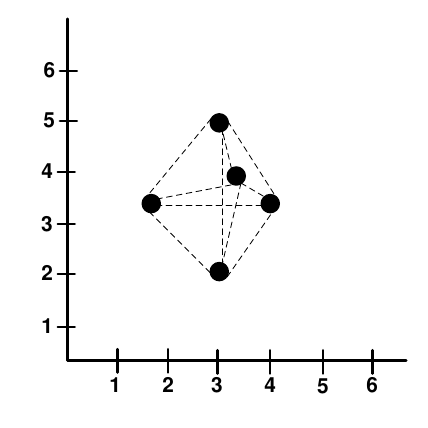}
                \caption{All-$\epsilon$-connected}
        \end{subfigure}%
        ~ 
        \begin{subfigure}[b]{0.23\textwidth}
                \includegraphics[width=\columnwidth]{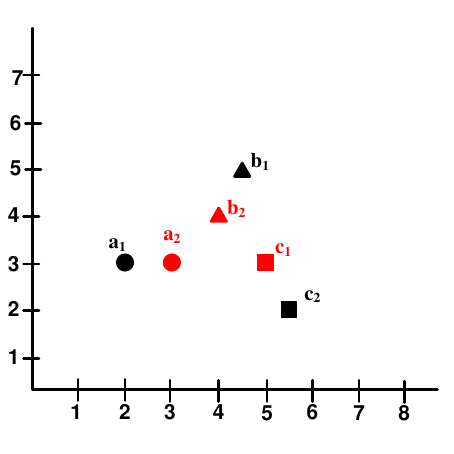}
                \caption{Input tuples in the 2D space}
        \end{subfigure}
        \caption{The notions of similarity within a group $\epsilon=3$}
        \label{fig:sgb-notion}
\end{figure}

In this section, we use distance as a measure to develop the notion of similarity among data elements that constitute a single group. Refer to Figure~\ref{fig:sgb-notion} for illustration. In Figure~\ref{fig:sgb-notion}a, the distance between all pairs of data elements satisfies the predefined similarity threshold $\epsilon=3$. The notable characteristic of this grouping is that elements are dense and form a clique-like group.
\begin{sloppypar}
\begin{definition}
\textbf{Group Compactness (Gpact)}. Group Compactness (or Gpact, for short) is the number of tuple
pairs $\langle t_i, t_j\rangle$ inside a group, say $G_m$, where the similarity predicate $\xi_{\delta,\epsilon}(t_i, t_j)$ is true. 
\end{definition}
\end{sloppypar}

\begin{sloppypar}
\begin{definition}
Given a set of tuples that form a group, say $G_{all}$.
$G_{all}$ is said to be \textbf{All-$\epsilon$-Connected} if, for all pairs $t_i$, $t_j \in G_{all}$,
the similarity predicate between $t_i$ and $t_j$ is true, i.e.,
\begin{equation*}
\forall t_i, t_i \in G_{all}, \xi_{\delta,\epsilon}(t_i, t_j)
\end{equation*}
\end{definition}
\end{sloppypar}

\begin{proposition}
Let $G_{all}$ be an All-$\epsilon$-Connected group that contains $k$ tuples. Then,
\begin{equation*}
Gpact(G_{all}) =  k*(k-1)/2.
\end{equation*}
\end{proposition}





\subsubsection{Semantics of Order-independent SGB Operators}


Given an input table, say $T$, of tuples, an SGB operator generates an output set of groups, say $G_s$. The various mappings between the tuples of $T$ and the groups $G_s$ determine the semantics of the various SGB operators.

\begin{sloppypar}
\textbf{SGB-All}: SGB-All divides tuples in $T$ into groups in $G_s$, such that each group in $G_s$ is All-$\epsilon$-Connected and is maximal. Notice that a Group $G_m$ is maximal when no group $G_m'$ exists such that $G_m \subset G_m'$ and $G_m'$ is All-$\epsilon$-Connected. Formally,
\end{sloppypar}

Refer to Figure~\ref{fig:sgb-notion}(b) for illustration. In the figure, tuples are modeled as points in the 2D space.
When $\epsilon=3$, applying the SGB-All semantics on the input tuples produces the three groups: $G_1=\left\lbrace a_1, a_2, b_2, c_1\right\rbrace$, $G_2=\left\lbrace b_1, a_2, b_2, c_1\right\rbrace$ and $G_3=\left\lbrace c_1, c_2, a_2, b_2\right\rbrace$. \\



\begin{sloppypar}
\begin{definition}
     \textbf{Overlap Set} $OSet$: A tuple is in the set $OSet$ iff the tuple belongs to more than one group in $(G_1, ..., G_s)$. 
    Formally, $Oset =\cup_{(i,j) \in s} (G_i \cap G_j )$ where $s>1$.
\end{definition}
Referring to the input tuples in Figure \ref{fig:sgb-notion}(b), the overlap set $OSet$ is comprised of $\left\lbrace a_2, b_2, c_1\right\rbrace$.
\end{sloppypar}

\subsubsection{Similarity Group-by ALL (SGB-All)}
The corresponding SQL syntax for the SGB-All operator is as follows:\\
{\ttfamily
\hspace*{1ex}\hspace{1ex}\hspace{1ex}\hspace{1ex}SELECT \textit{column}, aggregate-func(\textit{column})\\
\hspace*{1ex}\hspace{1ex}\hspace{1ex}\hspace{1ex}FROM \textit{table-name}
\newline
\hspace*{1ex}\hspace{1ex}\hspace{1ex}\hspace{1ex}WHERE \textit{condition}
\newline
\hspace*{1ex}\hspace{1ex}\hspace{1ex}\hspace{1ex}\textbf{GROUP BY} \textit{column} \textbf{DISTANCE-TO-ALL} [\textit{$L_2$} $ \mid $
\newline
\hspace*{1ex}\hspace{1ex}\hspace{1ex}\hspace{1ex}\textit{LINF}] \textbf{WITHIN} \textit{eps}
\newline
\hspace*{1ex}\hspace{1ex}\hspace{1ex}\hspace{1ex}\textbf{ON-OVERLAP[} \textit{DUPLICATE} $\mid $ \textit{ELIMINATE} $ \mid $ \\
\hspace*{1ex}\hspace{1ex}\hspace{1ex}\hspace{1ex}\textit{NEW-GROUP}]
}

SGB-All uses the following clauses to realize similarity-based grouping:
\begin{itemize}

\item The DISTANCE-TO-ALL similarity predicate specifies the metric space distance function to be applied when deciding group memberships.
\begin{itemize}
\item $L_{2}$: Euclidean distance
\item LINF: the Maximum infinity distance
\end{itemize}
\item ON-OVERLAP is an arbitration clause to be used in the situation when a tuple is within $eps$ distance from more than one group.
\end{itemize}

\begin{figure}
\centering
\includegraphics[width=3in,height=2in]{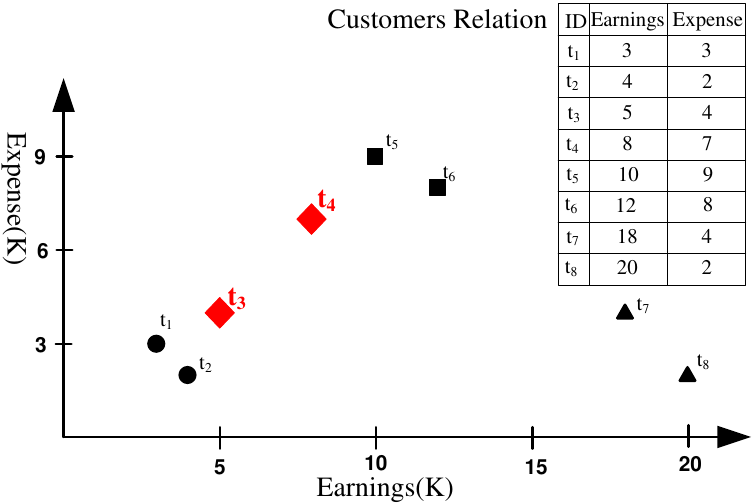}
\caption{Similarity-based Grouping $\epsilon=6$}
\label{figure:sgb-order-ind-v3}
\end{figure}

The following are the possible arbitration actions of the ON-OVERLAP clause:
\begin{itemize}
\item \textbf{DUPLICATE}: inserts an input tuple, say $t$, into all the groups that $t$ belongs to. 
Given an input table, say  $T$, the ON-OVERLAP DUPLICATE option is interpreted as inserting all the tuples of $OSet$ into their corresponding
overlapped groups from $G_s$. 
The ON-OVERLAP DUPLICATE option maps to the problem of listing maximal cliques in an undirected graph.
Analysis of the ON-OVERLAP DUPLICATE option is presented in Section~\ref{sgb-alg}.
\item \textbf{ELIMINATE}: discards the tuples that overlap multiple groups. Given an input table, say  $T$, the ON-OVERLAP ELIMINATE option is interpreted as deleting all the tuples of $OSet$ from all overlapped groups.
\item \textbf{NEW-GROUP}: inserts the overlapped data elements into a new group. Given an input table, say $T$, the ON-OVERLAP NEW-GROUP option is interpreted as creating new group sets for $OSet$ until $OSet$ is empty. Analysis of this option is presented in Section~\ref{sgb-alg}.
\end{itemize}

To illustrate, the following query example performs the aggregate operations $max$, $min$ and $count$ on the groups formed by SGB-All on attributes $Earnings$ and $Expense$ from relation $Customers$ given in Figure~\ref{figure:sgb-order-ind-v3}.\\

\noindent
{\ttfamily
\hspace*{1ex}\hspace{1ex}\hspace{1ex}\hspace{1ex}SELECT  min(\textit{earnings}), max(\textit{expense}), count(*)
\newline
\hspace*{1ex}\hspace{1ex}\hspace{1ex}\hspace{1ex}FROM \textit{customer}
\newline
\hspace*{1ex}\hspace{1ex}\hspace{1ex}\hspace{1ex}\textbf{GROUP BY} \textit{earnings}, \textit{expense} \textbf{DISTANCE-TO-ALL L2}
\newline
\hspace*{1ex}\hspace{1ex}\hspace{1ex}\hspace{1ex}\textbf{WITHIN} \textit{6}
\newline
\newline
}
Referring to the input table in Figure \ref{figure:sgb-order-ind-v3}, the following groups satisfy the above SGB-All predicates: $G_1=\left\lbrace t_1, t_2, t_3 \right\rbrace$, $G_2=\left\lbrace t_4, t_5, t_6 \right\rbrace$ and $G_3=\left\lbrace t_7, t_8 \right\rbrace$. However, $t_3$ is also within $eps$ from $t_4$ in $G_2$. Consequently, an arbitration  ON-OVERLAP clause is necessary. We consider the three possible semantics below for illustration.

If we have an ON-OVERLAP DUPLICATE clause, since $t_3$ is within $eps$ from $t_4$ but is not within $eps$ from $t_5$ or $t_6$, $t_3$
cannot be added to $G_2$. Similarly, $t_4$ cannot be added into $G_1$ since $t_4$ is not within $\epsilon$ from $t_1$ and $t_2$ in $G_1$.
Therefore, a new group is created, say $G_4=\left\lbrace t_3, t_4 \right\rbrace$ and is produced in addition to the three other groups. Therefore, the answer to the query in the case of
ON-OVERLAP DUPLICATE is: $\left\lbrace \langle 3, 4, 3 \rangle , \langle 8, 9, 3 \rangle, \langle 18, 4, 2 \rangle , \langle 5, 7, 2 \rangle \right\rbrace$.

\begin{sloppypar}
If we have an ON-OVERLAP ELIMINATE clause, the overlapped tuples $t_3$ and $t_4$ get dropped. Therefore, the resulting groups are $G_1=\left\lbrace t_1, t_2 \right\rbrace$, $G_2=\left\lbrace t_5, t_6 \right\rbrace$ and $G_3=\left\lbrace t_7, t_8 \right\rbrace$ and the query output is  $\left\lbrace \langle 3, 3, 2 \rangle , \langle 10, 9, 2 \rangle, \langle 18, 4, 2 \rangle \right\rbrace$.
\end{sloppypar}

If we have an ON-OVERLAP NEW-GROUP clause, the overlapped tuples are inserted into newly created group sets until overlap sets are empty. The result groups are $G_1=\left\lbrace t_1, t_2 \right\rbrace$, $G_2=\left\lbrace  t_5, t_6 \right\rbrace$, $G_3=\left\lbrace t_7, t_8 \right\rbrace$ and $G_4=\left\lbrace t_3, t_4 \right\rbrace$, and the query output will be  $\left\lbrace \langle 3, 3, 2 \rangle , \langle 10, 9, 2 \rangle, \langle 18, 4, 2 \rangle, \langle 5, 7, 2 \rangle \right\rbrace$.
Notice that this output is un-ambiguous in the sense that it should be independent of the order in which the input tuples are processed.

\section{Applications and Graph Representation of SGB}
\label{sec-app}
\subsection{Applications}
This section presents several application scenarios that demonstrate the practical use of the various SGB semantics. 

\begin{sloppypar}
\textbf{Highly Correlated Stock Analysis:}
To analyze stocks over time, \cite{Boginski03onstructural} converts the stock market datasets to a graph by mapping one stock into a vertex, then connecting two vertices iff their correlation coefficient is no smaller than a user-specified threshold. Stocks in the graph that form a clique are very meaningful from the application's point of view, as this implies that the prices of these stocks are contained in the same group that usually evolve synchronously over time, and a change of one stock price can be used to predict a similar change for the prices of all the other stocks in the same group. 
The proposed SGB-All semantics naturally construct these cliques, and each tuple in the same group stands the clique properties. Thus, people can analyze the highly correlated stock profiles using the SGB-All operator.
\end{sloppypar}

\begin{sloppypar}
\textbf{Geo-social network analysis:}
When users check-in or share photos in social networks,
the server records the users' spatial locations, e.g., their latitude and longitude. This geo-spatial information may reflect common behavior of people in a group. It may also help in marketing. For example, people who happen to be watching a concert or a movie, or go for shopping and have similar geo-locations, these shoppers stand a higher possibility to share the same interest or preference with products. In this scenario, SGB-All queries can identify groups where people are highly correlated given their spatial locations. If one individual shows interest in a product, e.g., a certain brand of T-shirts, using SGB-All, 
In this case, the social network can recommend targetted advertisements of certain products to the people in the same group as they are likely to show similar interest.
\end{sloppypar}


\subsection{Graph Representation of SGB}
\begin{figure}
\centering
\includegraphics[width=3.3in,height=2.4in]{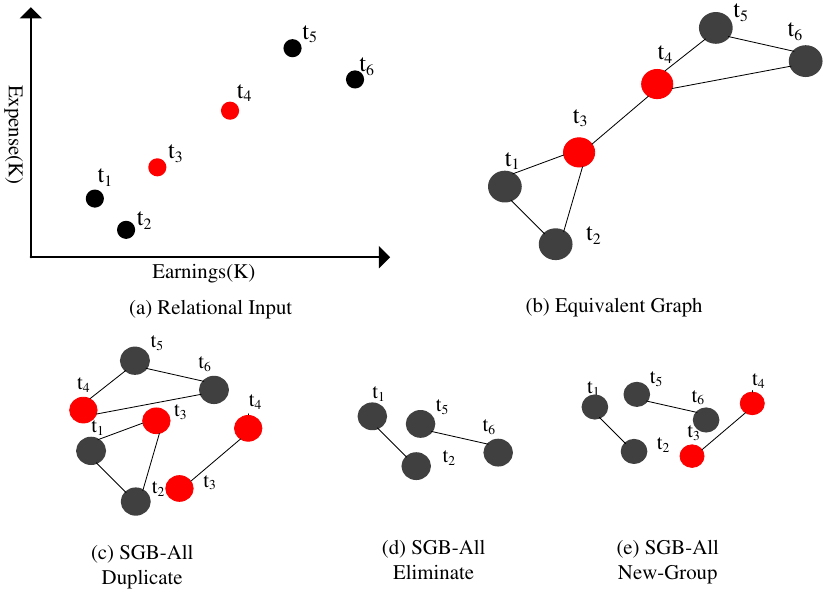}
\caption{Graph representations of the various SGB semantics.}
\label{figure:sgb-graph}
\end{figure}

\begin{sloppypar}
We use a graph to represent the semantics of each of the SGB operators. Refer to Figure~\ref{figure:sgb-graph}a that gives tuples in the 2D space and their corresponding equivalent graph representation in
Figure~\ref{figure:sgb-graph}b. The relation is mapped into the equivalent graph as follows.
Each input tuple is mapped to a vertex. An edge connects a pair of tuples, say $t_i$ and $t_j$, whenever
the similarity predicate $\xi_{\delta,\epsilon}(t_i, t_j)$ is true.
\end{sloppypar}

\begin{sloppypar}
Referring to Figures~\ref{figure:sgb-graph}c-e, the semantics of SGB-All Duplicate finds all maximal cliques in the graph.
The SGB-All Eliminate discards all overlapped tuples and finds the cliques that are not overlapped among more than one
group. The cliques in SGB-All Eliminate case are not maximal because the overlapped tuples are dropped.
The SGB-All New-Group forms new group(s) from overlapped tuples that translate to finding the overlapped cliques.
\end{sloppypar}


Notice that it is possible to extract the tuples and construct the corresponding graph outside of the database. Then, by using any graph-based processing tool, we can find and produce the cliques as output. However, for the reasons stated in the introduction section, mainly, to avoid the impedence mismatch of having to export data out of the database and then import the results back into the database for further processing, we want to perform this clique detection operation and the SGB-ALL operation from within the database system.
As we scan the input tuples using a table-scan operator, we want to partially construct the corresponding graph on the fly as well as detect the formed cliques. Notice that as we see more tuples, the existing cliques are likely to grow while new cliques will form or split from exising ones.



\section{Order-inde\-pen\-dent Algorithm \\for the SGB Operator}
\label{sgb-alg}
In this section, we present an order-indepedent algorithm for the SGB-All operator that handles all the three semantic varieties presented earlier in the paper. Then, we 
analyze the complexity of the algorithm and prove its order-independence. 

\subsection{Outline of the Algorithm}

The algorithm has two main building blocks. 
The first building block decides on which groups to join for a given input tuple. 
The second building block handles the various overlap semantics for SGB. 

Let $G_s$ be the set of existing groups constructed so far by the Algorithm.
When a new tuple, say $t_i$, is being processed, we find the set of groups, 
say $G_m \subseteq G_s$, that $t_i$ can join. 
Notice that $G_s$ is initially empty. 
In this case, the first tuple to be processed will form a new group. 
Based on the various SGB-All semantics, Tuple $t_i$ can join Group $g \in G_m$ iff $\forall t_j \in g$, the distance between $t_i$ and $t_j$ is smaller than the threshold $\epsilon$. 
To identify which groups to join, the first building block, $GQ(t_i, \epsilon, \delta)$, performs the above test and returns all the groups $G_m \subseteq G_s$  that $t_i$ will join. 

The straightforward way to evaluate $GQ(t_i, \epsilon, \delta )$ is to perform distance comparisons between $t_i$ and each of the processed tuples that are members of the existing groups in $G_s$ found thus far. 
Refer to Figure \ref{fig:GroupBound}(a) for illustration, where $L_\infty$ is assumed to be the underlying distance metric. 
In the figure, the following six tuples $t_1 \cdots t_6$ have already been processed, and two groups $g_1$ and $g_2$ have already been formed. 
When it is time to process Tuple $t_7$,
this involves performing distance comparisons between $t_7$ and each of the tuples ($t_1, t_2, t_3$) that form Group $g_1$, and then performing distance comparisons between $t_7$ and each of the tuples ($t_4, t_5, t_6$) that form $g_2$. 
$GQ(t_i, \epsilon, \delta)$ performs this operation in a more efficient way as described later in this section.




\begin{sloppypar}
The second building block handles partial group overlaps. Instead of joining existing groups, the input tuple $t_i$ may have partial overlaps with individual members of existing groups (not entire groups). In this case, $t_i$ can form new cliques with these tuples.
We start by excluding all the groups that $t_i$ has already joined after executing the first building block, i.e., excluding the output of $GQ(t_i, \epsilon, \delta)$. Thus, for all the remaining groups, i.e.,  
$\hat{G} \in G_s$ and $ \hat{G} \notin GQ(t_i, \epsilon, \delta)$, 
the second building block, $OQ(t_i, \epsilon,\delta)$, retrieves tuples from within $\hat{G}$ and decides if new cliques needs to be formed. These are the tuples that are enclosed within $\epsilon$ from Tuple $t_i$. 
Refer to Figure  \ref{fig:GroupBound}(a) and assume an $L_\infty$ distance function. For tuple $t_7$, there are no tuples that are within $\epsilon$ from Tuple $t_7$ (the red square in the figure). 
\end{sloppypar}

\begin{sloppypar}
Algorithm 1 gives the pseudo-code of the SGB-All operator using the straightforward all-pairs comparisons scheme described above. Line 1 iterates over all tuples, and for each tuple, searches all the existing groups in Set $G_s$, Function $GroupQuery()$ (Line 2) checks whether the new tuple $t_i$ can join one of the existing groups, say $g_m$, and it takes $k$ distance comparisons to decide if $t_i$ is within $\epsilon$ from each member tuple of $g_m$, where $k$ is expected number of tuples in $g_m$. If one of these distances, say between $t_i$ and $t_j$ is bigger than $\epsilon$, $t_i$ fails to join Group $g_m$ and searches the next group. Next, Function $OverlapQuery$ determines the groups with partial overlap based on the all-pairs distance comparison (Line 3). Later, the post-processing step handles these overlap tuples among the various groups that they belong to(Line 4-12). More detail about this step is given below. 
Given the two \textit{for-loop} iterations of $GroupQuery$ and $OverlapQuery$, the total complexity of this straightforward approach is bound by $O(n^2)$. We need to reduce the computation costs of both the group join query $GQ(t_i, \epsilon, \delta)$ and the overlap query $OQ(t_i, \epsilon, \delta)$.
\end{sloppypar}

\begin{algorithm}
\KwIn{Data Tuples in Relation T, Similarity threshold $\epsilon$, distance function $\delta$, overlap option $OL$}
\KwOut{Set of groups $G_s$}
\For{ each data element $t_{i}$ in $T$}
{

$GQ(t_i,\epsilon, \delta) \gets GroupQuery(t_i, G_s, \epsilon, \delta)$\\
$OQ(t_i, \epsilon, \delta)  \gets OveralpQuery(t_i,G_s,\epsilon, \delta)$\\

\If{ $OL$ is Duplicate}
{

     find all groups, $G_{new}$ from $G_s$ that $t_i$ forms a clique with\\
     duplicate $t_i$ into each of these groups\\

}\ElseIf{ $OL$ is Eliminate}
{
   remove the tuples that overlap multiple groups from the tuples in the groups in $G_s$\\
}\ElseIf{$OL$ is Form-New}
{
   find all the tuples that are within $\epsilon$ from $t_i$\\
   save these overlap tuples for next round’s processing\\
}

\If{$GQ(t_i,\epsilon,\delta)$ is empty and $OQ(t_i, \epsilon, \delta)$ is empty}
{
    Form a new group $g_{new}$ and insert $g_{new}$ into $G_{s}$ \\
}

}
\caption{Skeleton Procedure for SGB-All}
\end{algorithm}

\subsection{Identifying Which Groups To Join}

\begin{sloppypar}
For an input tuple, say $t_i$, and the groups identified so far, Procedure $GQ(t_i, \epsilon, \delta)$ identifies all the groups that $t_i$ can join, i.e., when $t_i$ forms a clique with all member tuples of these groups. Because this procedure is costly, the target is to reduce the cost of $GQ(t_i, \epsilon, \delta)$. 
The main idea is to construct a border around individual groups so that for each group, say $g_m$, we can test whether the new tuple $t_i$ forms a clique with member tuples of $g_m$ by only checking $t_i$ against the border of $g_m$. This helps avoid the costly checking of the distance between $t_i$ and each tuple inside $g_m$.
The shape of the border of a group depends on the underlying distance metric. For the  $L_2$ distance metric, we propose to use an $\epsilon$-convex hull border around each group, while for the $L_\infty$ distance metric, we propose to use an $\epsilon$-rectangular border as explained below. 
\end{sloppypar}

\begin{sloppypar}
\textbf{Group borders for the $L_2$ distance metric.}
Let $g_m$ be a group that contains multiple points that are all within a Distance $\epsilon$ from each other, i.e., that form a clique.
For the $L_2$ distance metric, the group border for Group $g_m$ is a convex hull that encompasses all the points in the group. Observe that the diameter $d$ of the convex hull, i.e., the largest distance between any two points on the convex hull, is less than $\epsilon$. 
To decide if Tuple $t_i$ is a member of $g_m$, i.e., is within Distance $\epsilon$ from all tuples contained in $g_m$, it is enough to check if $t_i$ is inside $g_m$'s enclosing convex hull.

Thus, to obtain $GQ(t_i, \epsilon, L_2)$, we iterate over all groups, and return all the groups $g$ such that $t_i$ is inside the $\epsilon$-convex-hull of $g$. as stated in Lemma \ref{Lemma:bound1} below.
\end{sloppypar}

\begin{lemma}
\label{Lemma:bound1}
Let  $g_m$ be a group where all of $g_m$'s points are within  $\epsilon$ from each other. When we construct the convex-hull for the points in $g_m$ (that we term the $\epsilon$-convex-hull for $g_m$), then if a new tuple $t_i$ is inside this $\epsilon$-convex-hull, then $g_m$ is contained in the $GQ(t_i, \epsilon, L_2)$.
\end{lemma}

\begin{figure}
        \centering
        \begin{subfigure}[b]{0.23\textwidth}
                \includegraphics[width=\columnwidth]{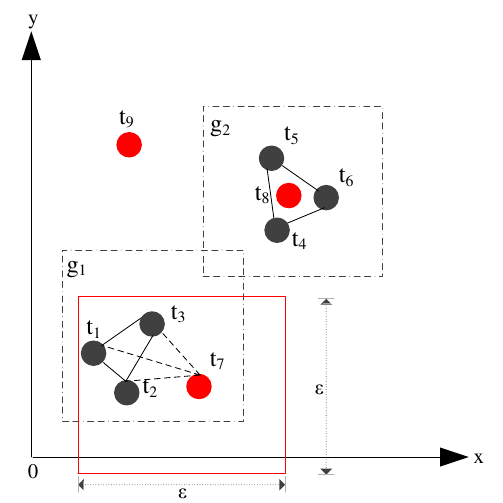}
                \caption{$\epsilon$-Group Bound}
        \end{subfigure}%
        ~ 
        \begin{subfigure}[b]{0.23\textwidth}
                \includegraphics[width=\columnwidth]{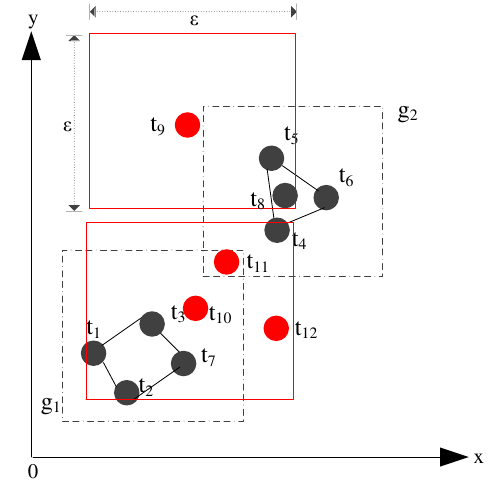}
                \caption{Handle overlap tuples}
        \end{subfigure}
        \caption{Illustration of SGB-All, tuples are processed in  order of their subscript, i.e., from $t_1$ to $t_{12}$.}
        \label{fig:GroupBound}
\end{figure}

\begin{proof}
The SGB-All semantics guarantees that $t_i \in g_m $ iff $\forall t_j \in g_m, j \neq i$, $\delta(t_i, t_j) \leq \epsilon $. Let the diameter, say $d$, of the convex hull be the maximum distance between any two points in the convex hull. Therefore,  $d$ should also be  less than  $\epsilon$. Consequently, if a new tuple $t_i$ is inside the convex hull of $g_m$, the distance from Tuple $t_i$ to any other tuple $t_j \in g_m $ is smaller than $\epsilon$. Otherwise, this will contradict the definition of the convex hull. Therefore, $t_i$ should also belong to $g_m$. Hence, $g_m$ is in $GQ(t_i, \epsilon, L_2)$.
\end{proof}

For example, consider the two triangles in Figure~\ref{fig:GroupBound}(a) that form the convex hulls for Groups $g_1$ and $g_2$. Tuple $t_8$ is located inside the convex hull for Group $g_2$ and is outside the convex hull for  Group $g_1$. Thus, $GQ(t_8, \epsilon, L_2)=\{g_2\}$. 
However, if a tuple is outside the convex hull, it can still form a clique with members of a group, and hence can still join the group.
For example, 
in Figure~\ref{fig:GroupBound}(a), Tuple $t_7$ is outside the convex hull for Group $g_1$. However, $t_7$ forms a clique with all points that compose Group $g_1$, and hence should join $g_1$, i.e., $GQ(t_7, \epsilon, L_2)=\{g_1\}$. 
Lemma~\ref{Lemma:ftdistance} handles this situation.

\begin{figure}
        \centering
        \begin{subfigure}[b]{0.23\textwidth}
                \includegraphics[width=\columnwidth]{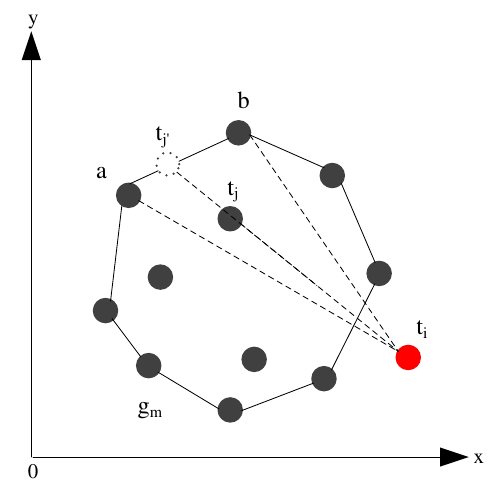}
                \caption{Illustration for Lemma~\ref{Lemma:ftdistance}.}
        \end{subfigure}%
        ~ 
        \begin{subfigure}[b]{0.23\textwidth}
                \includegraphics[width=\columnwidth]{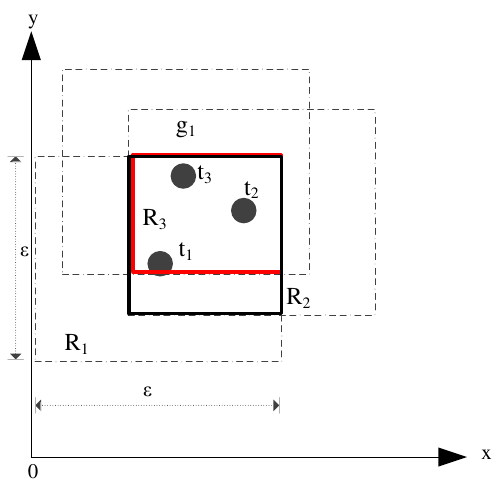}
                \caption{Illustration of how bounds are maintained according to Lemma~\ref{Lemma_groupdbound2}.}
        \end{subfigure}
        \caption{Illustration of the various group bounds for SGB-All. (a) A convex polygon in the case of the $L_2$ distance metric, and (b) A rectangle in the case of the $L_\infty$ distance metric. }
        \label{fig-Lemma}
\end{figure}

\begin{lemma}
\label{Lemma:ftdistance}
Let $g_m$ be a group whose elements are within Distance $\epsilon$ from each other. Let $t_i$ be a tuple outside the convex hull of $g_m$ such that the distance from $t_i$ to the farthest point in the convex hull is less than or equal to $\epsilon$. Then, $g_m$ is contained in $GQ(t_i, \epsilon, L_2)$.
\end{lemma}

\begin{sloppypar}
\begin{proof}
Assume that Group $g_m$ has $k$ points and that a tuple, say $t_j \in g_m$, is the point with farthest distance from Tuple $t_i$ among all the points of $g_m$. Tuple $t_j$ has to be one of the corner tuples of the convex hull of $g_m$ and cannot be an internal tuple in $g_m$. This can be shown by contradiction. Suppose that $t_j$ is an internal point of $g_m$ and the distance between $t_j$ and $t_i$ is the farthest compared to all the other tuples of $g_m$. Then, we can extend the line from Tuple $t_i$ to Tuple $t_j$, and this line would intersect one of the border lines of the convex hull, e.g., a line connecting between two points, say $t_a$ and $t_b$, on the border of the convex hull of $g_m$. Refer to Figure~\ref{fig-Lemma}a for illustration. Let the intersection point be $t_j'$. Naturally, $\delta(t_i, t_j')>\delta(t_i, t_j)$. In addition, based on simple triangular properties, one of the following enequalities has to be true: $\delta(t_i, t_a)>\delta(t_i, t_j')$ or  $\delta(t_i, t_b)>\delta(t_i, t_j')$. As a result, $\delta(t_i, t_j)$ is not the farthest distance and this contradicts the assumption. Therefore, if the distance from Tuple $t_i$ to the farthest point in Group $G_m$ (which has to be one of the points in the convex hull for $g_m$) is bigger than $\epsilon$, than $t_i$ cannot join in Group $g_m$.
\end{proof}
\end{sloppypar}

In Figure~\ref{fig:GroupBound}(a), Tuple $t_1$ has the farthest distance to $t_7$ from among all the tuples in Group $g_1$, where  this distance is smaller than $\epsilon$. Then, $GQ(t_7, \epsilon, L_2)=(g_1)$. More generally, a Group, say $g_m$, is contained in $GQ(t_i, \epsilon, L_2)$ if $t_i$ is inside the $\epsilon$-convex-hull bound of $g_m$, or if the longest distance from $t_i$ to the points in the convex hull of $g_m$ is not larger than $\epsilon$. Notice that the $\epsilon$-convex-hull bound of Group $g_m$ needs to be updated each time a new tuple joins the group. The worst case takes place when all the points on the convex hull of $g_m$ are on the circumferance of a circle with diameter $\epsilon$. For each Group, say $g_m$, that has $k$ tuples, the expected average number of convex points is $log(k)$~\cite{NEED-Reference}. Thus, with Lemmas~\ref{Lemma:bound1} and~\ref{Lemma:ftdistance}, we can limit the distance comparisons inside any one group.
Thus, the average case of running time to judge a new tuple inside one convex hull is $log(log(k))$.

\textbf{Group bound for the $L_\infty$ distance}:
In the case of the $L_\infty$ distance metric, a rectangle  replaces the $\epsilon$-convex hull of a group. A tuple $t_i$'s $\epsilon$-region is a square with Length $2*\epsilon$ centered at $t_i$. A tuple, say $t_j$, is within an $L_\infty$ Distance $\epsilon$ from $t_i$ iff $t_j$ is inside $t_i$'s  $\epsilon$-region. When $t_j$ joins the group containing $t_i$, we shrink the group's $\epsilon$-region to maintain the following invariant: A tuple, say $t$, is within an $L_\infty$ Distance $\epsilon$ from all members of a group, say $g$, iff $t$ is inside the group's $\epsilon$-region. Refer to Figure~\ref{fig-Lemma}b for illustration. Initially, when $t_1$ is inserted, a square centered at $t_1$ with side $2\epsilon$ forms the $\epsilon$-region for $t_1$, i.e., any new tuple that overlaps this region will be with $\epsilon$ from $t_1$ in the $L_\infty$ distance metric. When Tuple $t_2$ is processed and is found to be within this $\epsilon$-region, the group is now updated to contain $t_1$ and $t_2$ and the $\epsilon$-region is shrunk to be the intersection of the $\epsilon$-regions for both of  $t_1$ and $t_2$, which is Rectangle R2 in the figure. When Tuple $t_3$ is processed and is found to be inside $R_2$, i.e., $t_3$ is within $\epsilon$ from all members of the group corresponding to $R_2$, i.e., the group containing Tuples $t_1$ and $t_2$. Once $t_3$ joins the group, the $\epsilon$-region for the group is shrunk to be equal to the intersection of $R_2$ and the $\epsilon$-region for $t_3$. Now, the $\epsilon$-region for the group \{$t_1$, $t_2$, $t_3$\} becomes Rectangle $R_3$.
The following lemma~\ref{Lemma_groupdbound2} summarizes the approach above to avoid distance computations to each member of a group while deciding on which groups ot join i.e., when evaluating $GQ(t_i, \epsilon, L_\infty)$.

\begin{lemma}
\label{Lemma_groupdbound2}
The $\epsilon$-rectangle bound of a group, say $g_m$, is one rectangle that is the intersection of all the $\epsilon$-regions of all the tuples $t_j \in g_m$. If a new tuple, say $t_i$, is inside $g_m$'s $\epsilon$-rectangle, then $\forall t_j \in g_m$, $\delta(t_i, t_j) \le \epsilon$, and group $g_m$ is contained in $GQ(t_i, \epsilon, L_\infty)$.
\end{lemma}

Refer to Figure~\ref{fig:GroupBound}(a) for further illustration. In the figure, it is assumed that the tuples are processed according to their subsript index, i.e., from $t_1$ to $t_{12}$ in this order. By the time Tuple $t_3$ is processed, the $\epsilon$-regions for Tuples $t_1$, $t_2$, and $t_3$ intersect to form the surrounding rectangle with a dotted line boundary in the lower-left of the figure that corresponds to the group bound for $g_1$. Similarly, by the time Tuple $t_6$ is processed, 
the $\epsilon$-regions for Tuples $t_4$, $t_5$, and $t_6$ intersect to form the surrounding rectangle with a dotted line boundary that corresponds to the group bound for $g_2$. When  Tuple $t_9$ is processed, it is found to be outside of the existing two  group bounds for $g_1$ and $g_2$. Hence,  $GQ(t_9, \epsilon, L_\infty)$ is empty and $t_9$ forms a new group by itself. This way, we only check whether $t_i$ is inside the existing group bounds, which is performed in constant time per group regardless of the number of tuples in each group.

\begin{sloppypar}
\textbf{Group bound by $\epsilon$-rectangle and $\epsilon$-convex hull together}
For the $L_2$ distance metric, the $\epsilon$-rectangle bound also can be applied to filter out the unnecessary distance computation with respect the $\epsilon$-convex hull of group. For example, i.e., Figure \ref{fig:GroupBound} (a), two triangles in this example are the convex hull group bound for Group $G_1$ and $G_2$, which are enclosed by the $\epsilon$-rectangle of each group, thus, we can find $t_9$ can not join in group $G_1$ and $G_2$ for $L_2$ distance metric early, because $t_9$ is outside of the groups' $\epsilon$-rectangle. This is illustrated in the developed Lemma \ref{Lemma_groupdbound3}.
\end{sloppypar}

\begin{sloppypar}
\begin{lemma}
\label{Lemma_groupdbound3}
Given a group $G_m$, $G_m$ is contained in the $GQ(t_i, \epsilon, L_2)$ by Lemma \ref{Lemma:bound1} or Lemma \ref{Lemma:ftdistance}, then tuple $t_i$ is certain enclose by the $G_m$'s $\epsilon$-rectangle bound. On the contrary, one tuple stay inside Group $G_m$'s $\epsilon$-rectangle bound, yet, $G_m$ might not be contained in $GQ(t_i, \epsilon, L_2)$.
\end{lemma}
\end{sloppypar}

\begin{sloppypar}
\begin{proof}
This proof is based on the distance metric properties,that is, $\delta(t_i, t_j)$ by the $L_2$ distance metric $\ge$ $\delta(t_i, t_j)$ by  $L_\infty$ distance metric because of triangle properties. Based on this distance metric properties, we omitted the rest of proof because this prove by mathematical induction is straightforward.
\end{proof}
\end{sloppypar}

\begin{sloppypar}
\textbf{Index of group bound} is used to avoid searching through all exist group sets $G_s$. Multi-dimensional indexing like R-tree index over exist group sets $G_s$'s $\epsilon$-rectangle bound can enable group join query much quickly. Given a tuple $t_i$ with $\epsilon$-region, the range query to go through index to find groups, which are contained by the $\epsilon$-region of tuple $t_i$. Take Figure \ref{fig:GroupBound} (a) as an example, suppose one index like R-tree for the exist groups $G_1$, $G_2$'s $\epsilon$-rectangle bound,  $G_1$ and $G_2$ are not contained in $GQ(t_9, \epsilon, L_2)$ by searching through from index root to leaf node.
\end{sloppypar}

\subsection{Overlap Query}

\begin{sloppypar}
For overlap tuples among groups, there are three types of cases to be considered.
\begin{itemize}
\item $OQ(t_i, \epsilon, \delta)$ is empty, and $GQ(t_i, \epsilon, \delta)$ is not empty and queried group sets size is bigger than one, then $t_i$ is a overlap tuple. Consider the example in Figure\ref{fig:GroupBound}(b), $GQ(t_{11}, \epsilon,L_\infty)$=($G_1, G_2$), and tuple $t_{11}$ is enclosed by the $\epsilon$-rectangle bound of Group $G_1$ and $G_2$. Thus, $t_{11}$ is the overlap tuple for Group $G_{1}$ and $G_2$.

\item $GQ(t_i, \epsilon, \delta)$ is empty, $OQ(t_i, \epsilon, \delta)$ is not empty, tuples of $OQ(t_i, \epsilon, \delta)$ are overlap sets. For example, tuple $t_9$ in Figure\ref{fig:GroupBound}(b), $GQ(t_{9}, \epsilon, L_\infty)$ is empty, but tuples ( $t_5$, $t_8$) are contained in the $\epsilon$-region of $t_9$. Thus, after building new group $G_3$ for tuple $t_9$, tuples ($t_5$, $t_8$) are overlap sets between Group $G_2$ and $G_3$.

\item $GQ(t_i,\epsilon, \delta)$ and $OQ(t_i, \epsilon, \delta)$ is not empty, tuples of $OQ(t_i, \epsilon, \delta)$ and $t_i$ are overlap sets. Revisit Figure\ref{fig:GroupBound}(b) as a example, in term of tuple $t_{10}$, group join query $GQ(t_{10}, \epsilon, L_\infty)$=$G_1$, and overlap query $OQ(t_{10}, \epsilon, L_\infty)$=$t_4$, thus, ($t_{10}$, $t_4$) are overlap sets for Group $G_1$ and $G_2$.
\end{itemize}
\end{sloppypar}

\begin{sloppypar}
Naively, overlap query $OQ(t_i, \epsilon, \delta)$ approach goes through all exist group sets $G_s$, then retrieval overlap tuples in $t_i$'s $\epsilon)$-region. To avoid searching all the exist groups, we try to find overlap group candidates by the developed group bound in Lemma \ref{Lemma_groupdbound2}. Take Figure\ref{fig:GroupBound}(b) as an example, the $\epsilon$-region of $t_9$ is denoted as the red square, the region intersects with Group $G_2$'s $\epsilon$-rectangle bound, while it does not have intersection with $G_1$'s. Thus, we can assure that overlap sets between tuple $t_9$ and Group $G_1$ is empty, and overlap between tuple $t_9$ and Group $G_2$ is not empty. Hence, $\epsilon$-region and group $\epsilon$-rectangle bound is a optimal way to filter out some redundant computation checking, as stated in Lemma \ref{Lemma_overlap} below.
\end{sloppypar}

\begin{sloppypar}
\begin{lemma}
\label{Lemma_overlap}
Given a group $G_m$, $G_m$ is not in $GQ(t_i, \epsilon, \delta)$. $G_m$ has overlap tuple $t_i$ if $\epsilon$-rectangle bound of Group $G_m$ intersects with $\epsilon)$-region of tuple $t_i$.
\end{lemma}
\end{sloppypar}

\begin{sloppypar}
Proof for the Lemma \ref{Lemma_overlap} is straightforward base on the definition of $\epsilon$-region and the group rectangle bound of Lemma \ref{Lemma_groupdbound2}, we omit the details. As a result, the built multi-dimensional index over the group rectangle bound enable us to find overlap groups candidates efficiently without going through all the exist group.In addition, different from the $GQ(t_i, \epsilon,\delta)$, overlap query need to get overlap tuples from overlap tuples, which are belong to the overlap group candidates. The straight approach would go through all tuples of group $G_m$, and the optimal way is building multi-dimensional index e.g., Quadtree ~\cite{Quadtree1974}  for tuples $t_j \in G_m$ to support $\epsilon$-region range query of tuple $t_i$. 

\subsection{Handle overlap}
Follow the SGB-All semantic, there are three options to handle the overlap tuples among groups. For overlap Eliminate, the overlap tuples are removed from output sets. For New-Group option, the overlap tuples are materiel and processed by the next round of SGB-All operation until the overlap tuple sets are empty. Look at Figure\ref{fig:GroupBound}(b) as example, when processing tuple $t_9$, group join query $GQ(t_9, \epsilon, L_\infty)$ is empty, and overlap query $OQ(t_9, \epsilon,L_\infty)$=($t_3$, $t_8$). Thus,($t_3$, $t_8$) are marked as removable from Group $G_2$, and are processed again in the next round.
\end{sloppypar}

\begin{sloppypar}
For overlap-Duplicate option, if overlap tuples are from different exist groups, we need to check the maximal of new built group. Take the Figure\ref{fig:GroupBound}(b) as example, after finish processing tuple $t_1$ to $t_{11}$ based on SGB-All overlap-Duplicate semantic, Group $G_1$ contain tuples ($t_1$,$t_2$,$t_3$,$t_7$,$t_{10}$,$t_{11}$), Group $G_2$=($t_4$,$t_5$,$t_6$,$t_8$,$t_{11}$), Group $G_3$=($t_5$,$t_8$,$t_9$). For the new coming tuple $t_{12}$, group join query $GQ(t_{12}, \epsilon, L_\infty)$ is empty, and overlap query $OQ(t_{12}, \epsilon,L_\infty)$ contain tuples ($t_{7}$, $t_{10}$, $t_{11}$) from $G_1$, and ($t_{11}$) from $G_2$. From the SGB-All semantic, tuple ($t_{7}$, $t_{10}$, $t_{11}$, $t_{12}$) can form a new group $G_{4}$, while tuples ($t_{11}$, $t_{12}$) can form other new group $G_5$. However, group candidate $G_5$ is contained by group candidate $G_4$, therefore, only one group $G_4$ is built. Substring matching is an effective way for maximal groups checking, because tuple sets of one clique can be represented as a order string, then, one group is contained by others iff its mapped string is a substring of others'. Take the group candiate $G_4$ and $G_5$ as example, Group candidate $G_4$ is mapped to string \{7,10,11,12\}, and $G_5$ is mapped to string\{11,12\}. String \{11,12\} is contianed by string\{7,10,11,12\}, thus Group candidate $G_5$ is not maximal. 
\end{sloppypar}

\subsection{SGB-All Query Cost Model}
\begin{sloppypar}
In this section, we develop cost models for SGB-All search to estimate the running time of each overlap option. We at first give the running time cost for group join query $GQ(t_i, \epsilon, \delta)$ and overlap query $OQ(t_i, \epsilon, \delta)$, then analysis the cost for different overlap option. Although we mainly discuss the 2D space, the discussion can be extended to higher dimension.
\end{sloppypar}

\begin{sloppypar}
In term of group join query $GQ(t_i, \epsilon)$, for a Group $G_m$ with $k$ points, the expected size of $\epsilon$-convex-hull is $h=log(k)$, and judge tuple $t_i$ inside this convex hull can be done in $O(log(h))$ time \cite{DBLP:Atallah86}. In addition, as Lemma \ref{Lemma:ftdistance}, we need to check distance from $t_i$ to farthest tuple of $\epsilon$-convex-hull whether is smaller than $\epsilon$, and this can be done via $log(h)$ time. As a result, we can reduce naive all pairs distance computation from $k$ to $O(log(log(k)))$. Furthermore, multi-dimension index like R-tree \cite{Procedding:Rtree} can be used to index group's $\epsilon$-rectangle, and this can improve groups searching iteration time from number of exist groups ,say $|G_s|$, to $O(log(|G_s|))$. As a result, the running time complexity of the $GQ(t_i, \epsilon, \delta)$ can be enhanced from $O(n^2)$ to $O(n*log(|G_s|)*log(log(k)))$.
\end{sloppypar}
\begin{sloppypar}
In term of overlap query $OQ(t_i, \epsilon, \delta)$, based on the index over Group $G_m$'s $\epsilon$ rectangle, we can find overlap group candidates via $log(|G_s|)$ time following Lemma \ref{Lemma_overlap}. In addition, retrieval  overlap tuples from Group overlap candidate $G_m$ can be done in $O(log(k))$ based on the multi-dimensional index for tuples of overlap group candidate. Finally, suppose overlap group candidates size is $log(|G_s|)$ in average, the total cost for querying the overlap tuples is $O(n*log(|G_s|)+n*log(|G_s|)*log(k))$.
\end{sloppypar}
\begin{sloppypar}
Overall, sum of group join query and overlap query cost is $O(n*log(|G_s|)*log(k))$. We can observe that the running time of SGB-All query is depend on the output number of groups $|G_s|$ and number of tuples of one Group, say $k$. Furthermore, both two factors are depend on input parameter $\epsilon$ and data overlap option. We analysis the upper bound of output groups number $|G_s|$  for different overlap option below.
\end{sloppypar}

\begin{sloppypar}
\begin{lemma}
Given output group sets $G$ under SGB-All+Eliminate and SGB-All-New-Group option, any two groups $G_m, G_m'$ of $G$ do not share any tuples, and the upper bound of output group sets size $|G_s|$ is $O(n)$.
\label{corollary: onetuplerule}
\end{lemma}
\end{sloppypar}

\begin{sloppypar}
\begin{proof}
This can be proved by contradiction. Assume that $t_i$ was shared by two groups $G_m$ and $G_m'$, then the $t_i \in Osets$ then $Osets=G_m \cap G_m'$ and $Osets$ is not empty. This contradicts the semantic of SGB-All+Eliminate and SGB-All+New-Group, where the overlap $Osets$ is always empty. Therefore, each tuple only belong to one group. As a result, the number of groups $|G_s|$ has an upper-bound $O(n)$.
\end{proof}
\end{sloppypar}

\begin{sloppypar}
\textbf{SGB-All+Eliminate}
Following lemma~\ref{corollary: onetuplerule}, the running time complexity for SGB-All+Eliminate via the group bound and multi-dimensional index approach is $O(n*log(n)*log(k))$ since $|G_s|\leqq n$.
\end{sloppypar}

\begin{sloppypar}
\textbf{SGB-All+New-Group} overlap tuples are recursively processed until the overlap sets are empty, then the running time complexity of New-Group is depend on the recursive depth. Let the recursive processing depth be $d$, and for each recursively process, number of tuples be $n_i$, output of group size be $|G_{si}|$ and average number of tuples of group be $k_i$, the total running time complexity is $O(d*n*log(n)*log(k))$, because  $\sum_{i=1}^{d} O(n_i*log(|G_{si}|)*log(k_i))< \sum_{i=1}^{d} O(n*log(|G_{s}|)*log(k)) < O(d*n*log(n)*log(k))$.
\end{sloppypar}

\begin{sloppypar}
\textbf{SGB-All+Duplicate is exponential delay}
Following the semantics of SGB-All+Duplicate, this operator finds the groups $G_s$ that are maximal clique sets.
For $G_m \in G_s$, assume that the average size of a group is $ave|G_m|=k$. Consider the $k$ tuple permutations problem for different groups, the number of groups would be $\Omega(k^2* \binom{n}{k})$. As $\epsilon$ increases, then each group cover more tuples, group sizes $|G|$ can be large, e.g., $k=n/2$. Thus, even if we build the group bound and index for each group, the complexity is inflect by the number of output groups, and this brings the processing delay to be super-polynomial. For example, variants of Moon and Mose algorithm \cite{Clique:Bomze99themaximum} can be shown to have worst-case running time $O(3^{n/3})$. Overall, for the duplicate option, the SGB-All query time is bound by the number of groups, and this delay can be exponential.
\end{sloppypar}

\section{Correctness of SGB Operators}
\label{sec-proofs}
\begin{sloppypar}
In this section, we prove that SGB-All are order-independent operators for the one- (1D) and
two-dimensional (2D) spaces by the proposed algorithm. 
As we introduced before, SGB-All query can be viewed as cliques listing from undirected graph. Therefore, SGB-All operator is a order independent semantic if we build graph first and apply cliques listing algorithm. In other words, the output cliques sets are identical because the correspond graph structure is fixed beforehand. However, follow our previous discussion, we want to scan tuple in minimal time and output groups without building graph beforehand. Thus, we need to prove the proposed $\epsilon$-group bound approach via Lemma  \ref{Lemma:bound1},\ref{Lemma:ftdistance} and \ref{Lemma_groupdbound2} do not change outputs. The proofs build on results obtained from two classes of intersection graphs, namely the interval
graph~\cite{EEEexample:FuGr65} for the 1D case
and the rectangle graph
(also referred to as the Boxicity graph~\cite{boxicity:2007}) for the 2D case.
\end{sloppypar}

\subsection{Order Independent of Interval Graph}

\begin{figure}
\centering
\includegraphics[width=3in,height=3in]{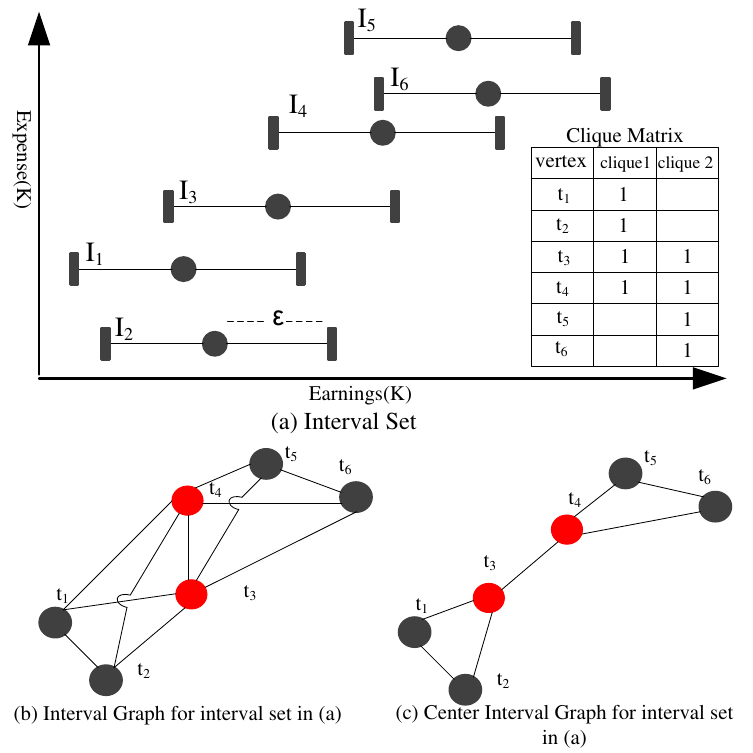}
\caption{(a) Intervals for the tuples in Figure~5, (b) their corresponding interval graph, and (c) their corresponding center interval graph.}
\label{fig:IntervalGraph}
\end{figure}

\begin{sloppypar}
\begin{definition}
Interval graph (\textit{IG}) is the intersection graph of a multiset of intervals on the real line.
\end{definition}
It has one vertex for each interval in the set, and connects two vertexes by an edge {\em iff}
their corresponding sets have nonempty intersection~\cite{EEEexample:FuGr65}. Figure~\ref{fig:IntervalGraph}a gives the
corresponding
interval set {$I1, I2,...I6$} for the Earnings attribute in Figure~\ref{figure:sgb-order-ind-v3}.
For instance, $I1$'s center point is Tuple $t_1$ and its corresponding interval varies
among (11-$\epsilon$,11+$\epsilon$), where $\epsilon=6$.
The interval set's corresponding interval graph is given in Figure~\ref{fig:IntervalGraph}b.
\end{sloppypar}

\begin{sloppypar}
\begin{lemma}~\cite{EEEexample:FuGr65}
A graph is an interval graph {\em iff} its corresponding clique vertex matrix exhibits the consecutive-ones property.
\label{corollary: consectuiveones}
\end{lemma}
\end{sloppypar}

\begin{sloppypar}
The clique matrix of an undirected graph is an incidence matrix having maximal cliques as rows and vertexes as columns~\cite{Intergraph:Terry1999,EEEexample:FuGr65}. Let $M$ be the clique matrix for a graph, say $G$.
$M$[i,j]=1 means that Vertex $i$ is one of the vertexes in Clique $j$.
Matrix $M$ is said to exhibit the consecutive-ones property in a given column if the rows can be permuted in a way
so that the ones can appear consecutively in that column. For example, in Figure~\ref{fig:IntervalGraph}(B),
we can observe that the clique matrix exhibits the
consecutive-ones property for the interval graph in the same figure. According to~\cite{EEEexample:FuGr65}, an
undirected graph G is an interval graph
if and only if the clique matrix of G has the consecutive ones property for columns.
\end{sloppypar}

\begin{sloppypar}
\begin{lemma}
Maximal clique sets in the interval graph are order-independent with respect to the order of the intervals.
\label{corollary:mclique_orderIndependent}
\end{lemma}
\end{sloppypar}

\begin{sloppypar}
\begin{proof}
For an interval graph,
its corresponding clique matrix, say $M_k$, differs according to the
order of introducing the intervals.
However, these different clique matrixes can always be reordered into one unique clique matrix as follows.
First, the interval order cannot change the intersection relations among the intervals.
So, the columns of all matrixes will always
be the same. Secondly, the rows of the clique matrix can be reordered to exhibit the consecutive -nes property.
Thus, the clique matrixes for different reshufflings of the order of the input intervals
will always lead to the same identical matrix by linear matrix transformations. As a result,
the maximal clique sets will be the same regardless of the order of processing of the intervals in the interval graph, and hence
are order-independent.
\end{proof}
\end{sloppypar}

\subsection{SGB-All+Duplicate for the 1D Space}
\begin{sloppypar}
Each interval have a center point. When two intervals intersect, it does not necessarily mean that the distance between their corresponding tuples, which are in the center, is smaller than $\epsilon$. As an alternative formation, consider the case when we connect vertexes in the interval graph iff an interval's center point is located inside another interval. This graph is referred to as a center interval graph (\textit{CIG}) as follow.
\end{sloppypar}

\begin{sloppypar}
\begin{definition}
$\epsilon$-Center interval graph ($\epsilon$-\textit{CIG}): let the interval diameter be $2*\epsilon$ and the center point be a tuple, say $t_i$, the related center interval graph is called $\epsilon$-\textit{CIG} iff vertexes in the interval graph have edges while an interval's center point is located inside another's interval.
\label{definition:CIG}
\end{definition}
\end{sloppypar}

\begin{sloppypar}
For example, Figure~\ref{fig:IntervalGraph}c gives the $\epsilon$-\textit{CIG} graph for the intervals in Figure~\ref{fig:IntervalGraph}a.
Notice that it is always the case that the center interval graph is always
a subgraph of the corresponding interval graph (refer to Figures~\ref{fig:IntervalGraph}b and~\ref{fig:IntervalGraph}c for illustration).
Observe that two vertexes that are connected by an edge in the $\epsilon$-\textit{CIG} implies that distance between the corresponding two tuples is within $\epsilon$.
\end{sloppypar}

\begin{sloppypar}
\begin{lemma}
$\epsilon$-\textit{CIG} is built via $\epsilon$-rectangle group bound in the one dimensional space, and SGB-All+Duplicate via $\epsilon$-rectangle bound approach finds the maximal clique sets in $\epsilon$-\textit{CIG}.
\label{corollary:SGB_MAXCLIQUE_SETS}
\end{lemma}
\end{sloppypar}

\begin{sloppypar}
\begin{proof}
Recall that in the SGB-All operator, tuples belong to one group $iff$ the similarity predicate
$\xi_{\delta,\epsilon}(t_i, t_j)$ is true. This grouping process in the 1D space is similar to the way
a $\epsilon$-\textit{CIG} graph is constructed by the $\epsilon$-rectangle group bound, because $\epsilon$-rectangle group bound stands the same distance criteria of interval graph.  Notice the $\epsilon$-rectangle group bound is an interval in one dimensional space. Meanwhile, each group $G_m \in G_s$ is an
all-$\epsilon$-Connected group, and the all-$\epsilon$-Connected group is a maximal clique when mapping
its tuples to the corresponding
graph. So, the maximal clique sets of the $\epsilon$-\textit{CIG} are the SGB-All+Duplicate via $\epsilon$-rectangle bound approach output groups $G_s$ in the 1D space.
\end{proof}
\end{sloppypar}

\begin{sloppypar}
\begin{lemma}
 \textbf{SGB-All+Duplicate is an order-independent operator in the 1D space via $\epsilon$-rectangle group bound.}
 \label{corollary:SGB_ORDERINDEPENDENT_ONEDIMENSION}
\end{lemma}
\end{sloppypar}

\begin{sloppypar}
\begin{proof}
From Lemma~\ref{corollary:mclique_orderIndependent}, the maximal clique sets in the interval graph are
order-independent. The $\epsilon$-\textit{CIG} is a subgraph of the corresponding interval graph,
the order-independent property is still true for building the $\epsilon$-\textit{CIG} subgraph via group bound Lemma \ref{corollary:SGB_MAXCLIQUE_SETS}.
This can be easily proven by contradiction (omitted here for brevity).
Therefore, we can conclude that SGB-All+Duplicate is order-independent if applying the $\epsilon$-bound in the 1D space based on
Lemma~\ref{corollary:SGB_MAXCLIQUE_SETS} as SGB-All+Duplicate via $\epsilon$-rectangle group bound approach finds the maximal clique sets in $\epsilon$-\textit{CIG}.
\end{proof}
\end{sloppypar}

\subsection{SGB-All+Duplicate for Higher Dimensional Spaces}

\begin{figure}
\centering
\includegraphics[width=3.3in,height=2.1in]{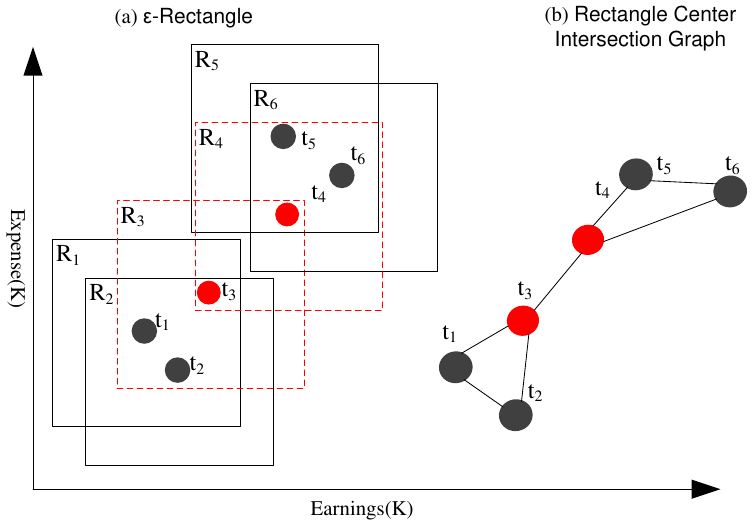}
\caption{(a)~$\epsilon$-Rectangles and (b)~$\epsilon$-\textit{RciG} for parts of the data in Figure~\ref{figure:sgb-order-ind-v3}.}
\label{fig:CriG}
\end{figure}

\begin{sloppypar}
We come to prove the SGB-All+Duplicate via group bound is still
order-independent in higher dimensions. We use the two-dimensional case for illustration. First, we introduce~\textit{RciG} to illustrate the $L_{\infty}$ distance between two tuples, say $t_i$ and
$t_j$. As we introduced before, for one tuple $t_i$, its $\epsilon$ range is a square with length $2*\epsilon$.
This implies that if one tuple, say $t_j$, is located inside other tuple $t_i$'s $\epsilon$-region $R_i$,
the $L_{\infty}$ distance between $t_i$ and $t_j$ is smaller than $\epsilon$. For example, in
Figure~\ref{fig:CriG}a,
Tuple $t_2$ is inside $t_1$'s square $R_1$, and the $L_{\infty}$ distance between $t_1 and t_2$ is smaller
than $\epsilon$. Thus, each vertex $v$ represents one square, and two vertexes have an edge between
if only if their center points $c$ are inside other's rectangle. Formally, we define \textit{RciG} as
follows.
\end{sloppypar}

\begin{sloppypar}
\begin{definition}
Rectangle center intersection Graph (\textit{RciG}), \textit{RciG} is a graph in which two vertexes, say $v_i$ and $v_j$, are connected by an  edge if only if
their center points stay in each other's corresponding rectangle. Specifically, if the rectangle's center point
is $t_i$ and the length of the rectangle is $2*\epsilon$, this rectangle is referred to as an
$\epsilon$-square for tuple $t_i$. The $\epsilon$-rectangle center intersection graph is termed  $\epsilon$-\textit{RciG}.
\end{definition}
\end{sloppypar}

\begin{sloppypar}
Figure~\ref{fig:CriG}a gives an example of $\epsilon$-rectangle intersection for parts of the data in
Figure~\ref{figure:sgb-order-ind-v3}, and Figure~\ref{fig:CriG}b is the corresponding $\epsilon$-\textit{RciG}. The $\epsilon$-rectangle intersection result is the group $\epsilon$-rectangle bound for the processed tuples.  Based on the $\epsilon$-rectangle, the maximal cliques in the $\epsilon$-\textit{RciG} graph are the all-$\epsilon$-Connected
groups produced as output by the SGB-All+Duplicate operator using the $L_\infty$ distance metric via group bound.
\end{sloppypar}

\begin{sloppypar}
To prove that $\epsilon$-\textit{RciG} shares relevant properties with $\epsilon$-\textit{CiG},
we can project the 2D rectangles in $\epsilon$-\textit{RciG} into the horizontal and vertical dimensions
separately, then the projections are the corresponding intervals in the two corresponding
$\epsilon$-\textit{CiG}s. For example, Rectangle $R_1$ in
Figure~\ref{fig:CriG} with center point $t_1(11,1)$, and its rectangle length is $2*\epsilon=12$. Then,
the projection of this rectangle in the horizontal dimension is $I1_{x}(11-6, 11+6)$ and in the vertical line is
$I1_y(1-6, 1+6)$.
\end{sloppypar}

\begin{lemma}
\label{corollary:CriGisIntersection}
 Two rectangles have intersection iff both of their projection intervals in the real lines intersect ,and $\epsilon$-\textit{RciG} is the intersection sets of $\epsilon$-\textit{CiG} in  one dimension.
\end{lemma}

Naturally,
the $\epsilon$-\textit{RciG} is the intersection sets of $\epsilon$-\textit{CiG}. From
Lemma~\ref{corollary:mclique_orderIndependent}, the maximal cliques sets in the interval graph via group bound are
order-independent. Thus, we use this conclusion to prove that the maximal cliques in the
$\epsilon$-\textit{RciG} are also order-independent by the following lemma.

\begin{sloppypar}
\begin{lemma}
\label{corollary:IntesectionOrderIndependent}
The intersection between two order-independent sets are also order-independent.
\end{lemma}
\end{sloppypar}

\begin{sloppypar}
\begin{proof}
We can prove this Lemma by contradiction. Suppose two input order sets $D_1$ and $D_2$, and their order-independent
output sets are $Gs_1$ and $Gs_2$, respectively. Assume that the intersection $x=(Gs_1 \cap Gs_2) \neq  null$,
and suppose that x is order-dependent in parts of $D_1$ and $D_2$. Let this part of the data in $D_1$
and $D_2$ be
$d_1'$ and $d_2'$, respectively. Because x is dependent on the order of $d_1'$ and $d_2'$, if we reorder $d_1'$ and $d_2'$, the
output x will change. When we have a different x, the order-independent outputs $Gs_1$ and $Gs_1$ would also be changed because x is always
the intersection of $Gs_1$ and $Gs_2$. This contradicts the assumption that $Gs_1$ and $Gs_2$ are order-independent on
$D_1$ and $D_2$. The reason is that from the order-independence definition, when we reorder part of the data in $D_1$ and $D_2$,
$Gs_1$ and $Gs_1$ still do not change. Thus, we prove that the intersection of two order-independent sets are also order-independent.
\end{proof}

\begin{lemma}
\label{corollary:L-infinitydistance-independent}
SGB-All+Duplicate operation via group $\epsilon$-rectangle bound is an order-independent operator for the $L_\infty$ distance in higher dimension.
\end{lemma}
\end{sloppypar}

\begin{sloppypar}
\begin{proof}
From Lemma~\ref{corollary:CriGisIntersection}, $\epsilon$-\textit{RciG} is the intersection of the two interval graphs, and the correspond maximal cliques in the interval graph built from group $\epsilon$-rectangle bound are order-independent. Thus, we can conclude that maximal cliques are order-independent in the 2D space according to the $L_\infty$ distance because the intersection of two order-independent sets are still order-independent according to  Lemma~\ref{corollary:IntesectionOrderIndependent}. As a result, SGB-All+Duplicate is an order-independent operator via $\epsilon$-group bound. More importantly, this proof can be extended to higher dimensions because we can project higher dimensional boxes to multiple lower-dimensional boxes, and
Lemma~\ref{corollary:CriGisIntersection} is always true for set intersection.
\end{proof}
\end{sloppypar}

\begin{sloppypar}
Next, we prove that SGB-All+Duplicate is order-independent according to the $L_2$ distance metric by the $\epsilon$-convex hull group bound. Tuple $t_i$ can be represented by a circle with Radius $\epsilon$, termed an $\epsilon$-circle. Similar to  the $\epsilon$-rectangle, any tuple, say $t_j$, that is inside $t_i$'s $\epsilon$-circle is within distance less than $\epsilon$ from $t_i$. Similar to the process for building the $\epsilon$-\textit{RciG} graph, we can build the $\epsilon$-circle intersection graph via the Lemma \ref{Lemma:bound1} and \ref{Lemma:ftdistance}, because $\epsilon$-convex hull group bound meets the distance criteria of building the $\epsilon$-circle intersection graph.
\end{sloppypar}

\begin{sloppypar}
\begin{lemma}
\label{corollary:l2-distance-independent}
The SGB-All+Duplicate operator via $\epsilon$-convex hull group bound is order-independent for the $L_2$ distance metric in higher dimensional spaces.
\end{lemma}
\end{sloppypar}

\begin{sloppypar}
\begin{proof}
For each tuple, say $t_i$, the $\epsilon$-circle inscribes the $\epsilon$-rectangle closely. Thus,
if one tuple, say $t_j$, is located inside $t_i$'s enclosing rectangle, then $t_j$ can still be outside of $t_i$'s circle. This means
that the $L_2$ distance between $t_i$ and $t_j$ can be larger than $\epsilon$. Thus, the $\epsilon$-\textit{RciG}
graph is the superset of the $\epsilon$-circle intersection graph. Because the maximal cliques in
$\epsilon$-\textit{RciG}, which are output from the SGB-All duplicate via $\epsilon$-rectangle group bound, are order-independent, then their subset is also order-independent. This can be proved by
contradiction. So, SGB-All+Duplicate via $\epsilon$-convex hull bound is still order-independent for $L_2$ distance metric.
\end{proof}
\end{sloppypar}

\begin{sloppypar}
\begin{conclusion}
\label{corollary:finalconclusion}
 \textbf{Operator SGB-All+Duplicate  is order-independent via group bound approach for the 1D and higher dimensions}
\end{conclusion}
\end{sloppypar}

\subsection{Order-independence of SGB-All+New-Group and SGB-All+Eliminate}

In the previous section, we have demonstrated that SGB-All+Duplicate is order-independent.
In this section, we prove that the other two overlap options, namely On Overlap Eliminate and and On Overlap New-Group are
also order-independent.

\begin{sloppypar}
\begin{proof}
For the output groups sets from the SGB-All+Duplicate operation, we can mark the overlap tuples $OSet$ by an
additional linear-time scan over $G_s$. Then, we can remove all the marked overlap tuples in $OSet$ from $G_s$ and put $OSet$ into a temporal tuple set $D_{tmp}$. We call this process is a one-time-iteration for SGB-All+New-Group. For tuples in $D_{tmp}$, we repeat
the one-time-iteration recursively until $D_{tmp}$ is empty. This is one possible procedure for the SGB-All+New-Group operator.
Notice that the algorithm of SGB-All+New-Group does not need to depend on SGB-All+Duplicate. For the one-time-iteration,We prefer to use Quadtree \cite{Quadtree} for its simplicity.
SGB-All+New-Group has identical output group sets on any order of $D_{tmp}$ because of Conclusion~\ref{corollary:finalconclusion}. Thus, SGB-All-New-Group is order-independent.
\end{proof}
\end{sloppypar}

Similarly, we can prove that SGB-All+Eliminate is also order-independent.

\begin{sloppypar}
\begin{proof}
We remove the overlapping tuples in $OSet$ from the output group sets $G_s$ of SGB-All+Duplicate. If SGB-All+Eliminate is
order-dependent, it means that the $OSet$ changes each time. This contradicts that SGB-All+Duplicate is order-independent because the overlap in the SGB-All+Duplicate not change.
\end{proof}
\end{sloppypar}
Finally, we conclude that
\begin{sloppypar}
\begin{conclusion}
 \textbf{SGB-All+New-Group and SGB-All+Eliminate via group bound are order-independent.}
\end{conclusion}
\end{sloppypar}






\section{Conclusion}
\label{sec-con}
\begin{sloppypar}
In this paper, we study the problem of order-independent similarity group by. This problem is motivated by the fact that traditional similarity group by operators do not necessarily produce identical results when the order of processing the input tables are altered. We define the class of order-independent similarity group by operators (SGB), then introduce the  SGB-All semantics to find the similar groups in the input sets. We provide the computational complexity costs of SGB-All, and demonstrate that useful variants of them can be implemented efficiently inside a database management system. Finally, using the notion of interval graphs, we prove that our newly developed SGB are order-independent by the efficient group bound approach. There are many interesting future directions that we plan to  explore. Different from SGB-All, we can propose SGB-Any semantic, which forms groups such that a data item, say O, belongs to a group, say G,
if and only if O is within a user-defined threshold from at least one other data item in G.  It will be interesting and useful to embed the operators developed in this paper inside a relational database engine. In fact, we have already started developing these SGB operators inside  PostgreSQL.
\end{sloppypar}


{
\small
\bibliographystyle{IEEEtran}

\bibliography{IEEEabrv,IEEEexample}

\begin{thebibliography}{10}
\providecommand{\url}[1]{#1}
\csname url@samestyle\endcsname
\providecommand{\newblock}{\relax}
\providecommand{\bibinfo}[2]{#2}
\providecommand{\BIBentrySTDinterwordspacing}{\spaceskip=0pt\relax}
\providecommand{\BIBentryALTinterwordstretchfactor}{4}
\providecommand{\BIBentryALTinterwordspacing}{\spaceskip=\fontdimen2\font plus
\BIBentryALTinterwordstretchfactor\fontdimen3\font minus
  \fontdimen4\font\relax}
\providecommand{\BIBforeignlanguage}[2]{{%
\expandafter\ifx\csname l@#1\endcsname\relax
\typeout{** WARNING: IEEEtran.bst: No hyphenation pattern has been}%
\typeout{** loaded for the language `#1'. Using the pattern for}%
\typeout{** the default language instead.}%
\else
\language=\csname l@#1\endcsname
\fi
#2}}
\providecommand{\BIBdecl}{\relax}
\BIBdecl

\bibitem{EEEexample:hall09}
M.~Hall, E.~Frank, G.~Holmes, B.~Pfahringer, P.~Reutemann, and I.~H. Witten,
  ``The weka data mining software: an update,'' \emph{SIGKDD Explor. Newsl.},
  vol.~11, no.~1, pp. 10--18, 2009.

\bibitem{IEEEexample:silva2009similarity}
Y.~N. Silva, W.~G. Aref, and M.~H. Ali, ``Similarity group-by,'' in \emph{Data
  Engineering, 2009. ICDE'09. IEEE 25th International Conference on}.\hskip 1em
  plus 0.5em minus 0.4em\relax IEEE, 2009, pp. 904--915.

\bibitem{VLDB:SilvaALPA13}
Y.~N. Silva, W.~G. Aref, P.-{\AA}. Larson, S.~Pearson, and M.~H. Ali,
  ``Similarity queries: their conceptual evaluation, transformations, and
  processing,'' \emph{VLDB J.}, vol.~22, no.~3, pp. 395--420, 2013.

\bibitem{IEEEexample:berkhin2006survey}
P.~Berkhin, ``A survey of clustering data mining techniques,'' in
  \emph{Grouping multidimensional data}.\hskip 1em plus 0.5em minus 0.4em\relax
  Springer, 2006, pp. 25--71.

\bibitem{IEEEexample:han2006data}
J.~Han, M.~Kamber, and J.~Pei, \emph{Data mining: concepts and
  techniques}.\hskip 1em plus 0.5em minus 0.4em\relax Morgan kaufmann, 2006.

\bibitem{IEEEexample:kanungo2002efficient}
T.~Kanungo, D.~M. Mount, N.~S. Netanyahu, C.~D. Piatko, R.~Silverman, and A.~Y.
  Wu, ``An efficient k-means clustering algorithm: Analysis and
  implementation,'' \emph{Pattern Analysis and Machine Intelligence, IEEE
  Transactions on}, vol.~24, no.~7, pp. 881--892, 2002.

\bibitem{EEEexample:zhang1996birch}
T.~Zhang, R.~Ramakrishnan, and M.~Livny, ``Birch: an efficient data clustering
  method for very large databases,'' in \emph{ACM SIGMOD Record}, vol.~25,
  no.~2.\hskip 1em plus 0.5em minus 0.4em\relax ACM, 1996, pp. 103--114.

\bibitem{IEEEexample:ester1996density}
M.~Ester, H.-P. Kriegel, J.~Sander, and X.~Xu, ``A density-based algorithm for
  discovering clusters in large spatial databases with noise.'' in \emph{KDD},
  vol.~96, 1996, pp. 226--231.

\bibitem{IEEEexample:anderberg1973cluster}
M.~R. Anderberg, ``Cluster analysis for applications,'' DTIC Document, Tech.
  Rep., 1973.

\bibitem{IEEEexample:mckee1999topics}
T.~A. McKee and F.~R. McMorris, \emph{Topics in intersection graph
  theory}.\hskip 1em plus 0.5em minus 0.4em\relax Siam, 1999, vol.~2.

\bibitem{EEEexample:Karp72}
R.~M. Karp, ``{Reducibility Among Combinatorial Problems},'' in
  \emph{Complexity of Computer Computations}, R.~E. Miller and J.~W. Thatcher,
  Eds.\hskip 1em plus 0.5em minus 0.4em\relax Plenum Press, 1972, pp. 85--103.

\bibitem{Clique:Bomze99themaximum}
I.~M. Bomze, M.~Budinich, P.~M. Pardalos, and M.~Pelillo, ``The maximum clique
  problem,'' in \emph{Handbook of Combinatorial Optimization}.\hskip 1em plus
  0.5em minus 0.4em\relax Kluwer Academic Publishers, 1999, pp. 1--74.

\bibitem{SIGMOD13:triangleProblem}
\BIBentryALTinterwordspacing
X.~Hu, Y.~Tao, and C.-W. Chung, ``Massive graph triangulation,'' in
  \emph{Proceedings of the 2013 ACM SIGMOD International Conference on
  Management of Data}, ser. SIGMOD '13.\hskip 1em plus 0.5em minus 0.4em\relax
  New York, NY, USA: ACM, 2013, pp. 325--336. [Online]. Available:
  \url{http://doi.acm.org/10.1145/2463676.2463704}
\BIBentrySTDinterwordspacing

\bibitem{Mdistance:Kruskal1964}
\BIBentryALTinterwordspacing
J.~Kruskal, ``\BIBforeignlanguage{English}{Multidimensional scaling by
  optimizing goodness of fit to a nonmetric hypothesis},''
  \emph{\BIBforeignlanguage{English}{Psychometrika}}, vol.~29, no.~1, pp.
  1--27, 1964. [Online]. Available: \url{http://dx.doi.org/10.1007/BF02289565}
\BIBentrySTDinterwordspacing

\bibitem{Boginski03onstructural}
V.~Boginski, S.~Butenko, and P.~M. Pardalos, ``On structural properties of the
  market graph,'' 2003.

\bibitem{Quadtree1974}
\BIBentryALTinterwordspacing
R.~Finkel and J.~Bentley, ``\BIBforeignlanguage{English}{Quad trees a data
  structure for retrieval on composite keys},'' vol.~4, no.~1.\hskip 1em plus
  0.5em minus 0.4em\relax Springer-Verlag, 1974, pp. 1--9. [Online]. Available:
  \url{http://dx.doi.org/10.1007/BF00288933}
\BIBentrySTDinterwordspacing

\bibitem{DBLP:Atallah86}
M.~J. Atallah, ``Computing the convex hull of line intersections,'' \emph{J.
  Algorithms}, vol.~7, no.~2, pp. 285--288, 1986.

\bibitem{Procedding:Rtree}
A.~Guttman, ``R-trees: A dynamic index structure for spatial searching,'' in
  \emph{international conference on management of data}.\hskip 1em plus 0.5em
  minus 0.4em\relax ACM, 1984, pp. 47--57.

\bibitem{EEEexample:FuGr65}
D.~R. Fulkerson and O.~A. Gross, ``{Incidence Matrices and Interval Graphs},''
  \emph{Pacific J. Math.}, vol.~15, pp. 835--855, 1965.

\bibitem{boxicity:2007}
\BIBentryALTinterwordspacing
L.~S. Chandran and N.~Sivadasan, ``Boxicity and treewidth,'' \emph{Journal of
  Combinatorial Theory, Series B}, vol.~97, no.~5, pp. 733 -- 744, 2007.
  [Online]. Available:
  \url{http://www.sciencedirect.com/science/article/pii/S0095895607000111}
\BIBentrySTDinterwordspacing

\bibitem{Intergraph:Terry1999}
\BIBentryALTinterwordspacing
T.~A. McKee and F.~R. McMorris, \emph{Topics in Intersection Graph
  Theory,Discrete Mathematics and Applications}.\hskip 1em plus 0.5em minus
  0.4em\relax SIAM, 1999. [Online]. Available:
  \url{http://epubs.siam.org/doi/book/10.1137/1.9780898719802}
\BIBentrySTDinterwordspacing

\end{thebibliography}
}





\end{document}